\documentclass[twoside,a4paper]{amsart}
\usepackage[printedin,extnum,nohyperref]{myamsart}
\ppnum{LEEDS--MATH--PURE--2004-18\\
}{arXiv: \href{http://arXiv.org/abs/cs.MS/0410044}{cs.MS/0410044}}
{2004}
\usepackage{noweb}
\usepackage{graphicx}
\nwfilename{parabolic0-bin.nw}
\makeatletter
\let\tab=&
\def\idxexample#1{\nwix@id@uses#1}
\makeatother

\def\nwlbrace{\textbf{\texttt{\char123}}}
\def\nwrbrace{\textbf{\texttt{\char125}}}
\newcommand{\CPP}{\texttt{C++}}
\newcommand{\NoWEB}{\texttt{noweb}}
\providecommand{\MetaPost}{\texttt{Meta}\-\texttt{Post}}
\providecommand{\GiNaC}{\textsf{GiNaC}}
\newif\iftth
\input{mydef}
\usepackage{xr-hyper}
\usepackage[hypertex]{hyperref}
\begin{document}
\def\LA{\begingroup\maybehbox\bgroup\setupmodname\Rm$\langle$}\def\RA{$\rangle$\egroup\endgroup}\providecommand{\MM}{\kern.5pt\raisebox{.4ex}{\begin{math}\scriptscriptstyle-\kern-1pt-\end{math}}\kern.5pt}\providecommand{\PP}{\kern.5pt\raisebox{.4ex}{\begin{math}\scriptscriptstyle+\kern-1pt+\end{math}}\kern.5pt}\def\commopen{/\begin{math}\ast\,\end{math}}\def\commclose{\,\begin{math}\ast\end{math}\kern-.5pt/}\def\begcomm{\begingroup\maybehbox\bgroup\setupmodname}\def\endcomm{\egroup\endgroup}\nwfilename{parabolic0-bin.nw}\nwbegindocs{1}%
\nwenddocs{}\nwbegindocs{2}%
\nwenddocs{}\nwbegindocs{3}%
\nwenddocs{}\nwbegindocs{4}%
\nwenddocs{}\nwbegindocs{5}%
\nwenddocs{}\nwbegindocs{6}%
\nwenddocs{}\nwbegindocs{7}%
\nwenddocs{}\nwbegindocs{8}%
\nwenddocs{}\nwbegindocs{9}%
\nwdocspar
\title{An Example of Clifford Algebras Calculations with \GiNaC}
\author[Vladimir V. Kisil]%
{\href{http://maths.leeds.ac.uk/~kisilv/}{Vladimir V. Kisil}}
\thanks{On  leave from Odessa University.}

\address{%
School of Mathematics\\
University of Leeds\\
Leeds LS2\,9JT\\
UK
}

\email{\href{mailto:kisilv@maths.leeds.ac.uk}{kisilv@maths.leeds.ac.uk}}

\urladdr{\href{http://maths.leeds.ac.uk/~kisilv/}%
{http://maths.leeds.ac.uk/\~{}kisilv/}}

\begin{abstract}
  This is an example of \CPP\ code of Clifford algebra calculations
  with the \GiNaC\ computer algebra system. This code makes both
  symbolic and numeric computations. It was used to produce
  illustrations for paper~\cite{Kisil04b,Kisil05a}.

  Described features of \GiNaC\ are already available at
  \texttt{PyGiNaC}~\cite{PyGiNaC} and due to course should propagate
  into other software like \texttt{GNU Octave}~\cite{Octave} and
  \texttt{gTybalt}~\cite{gTybalt} which use \GiNaC\ library as their
  back-end.

\end{abstract}
\maketitle

\tableofcontents

\section{Introduction}
This example of Clifford algebras calculations uses \GiNaC\ library
\cite{GiNaC}, which includes a support for generic Clifford algebra
starting from version~1.3.0.  Both symbolic and numeric calculation
are possible and can be blended with other functions of \GiNaC.
Described features of \GiNaC\ are already available at
\texttt{PyGiNaC}~\cite{PyGiNaC} and due to course should propagate
into other software like \texttt{GNU Octave}~\cite{Octave} and
\texttt{gTybalt}~\cite{gTybalt} which use \GiNaC\ library as their
back-end.

We bind our \CPP-code with
documentation using \NoWEB~\cite{NoWEB}
within \emph{the literate programming}
concept~\cite{LitProgFAQ}. Our program makes output of two
types: some results are typed on screen for information only and
the majority of calculated data  are stored in files which are lately
incorporate by \MetaPost~\cite{MetaPost} to produce PostScript
graphics for the paper~\cite{Kisil04b,Kisil05a}. Since this code can be treated
as software we are pleased to acknowledge that it is subject to GNU
General Public License~\cite{GNUGPL}.

\GiNaC\ allows to use a generic Clifford algebra, i.e. \(2^n\)
dimensional algebra with generators \(e_k\) satisfying the identities
\(e_ie_j+e_je_i=B(i,j)+B(j,i)\) for some (\emph{metric}) \(B(i,j)\),
which may be non-symmetric~\cite{Fauser96,FauserAblamowicz00a} and contain symbolic
entries.
Such generators are created by the function
\begin{webcode}{\Tt{}\Rm{}{\bf{}ex} {\it{}clifford\_unit}({\bf{}const} {\bf{}ex} & {\it{}mu}, {\bf{}const} {\bf{}ex} & {\it{}metr}, {\bf{}unsigned} {\bf{}char} {\it{}rl} = 0, {\bf{}bool} {\it{}anticommuting} = {\bf{}false});\nwendquote}
\end{webcode}
where {\Tt{}\Rm{}{\it{}mu}\nwendquote} should be {\Tt{}\Rm{}{\bf{}varidx}\nwendquote} class object indexing the
generators, an index {\Tt{}\Rm{}{\it{}mu}\nwendquote} with a \emph{numeric} value may be of type
{\Tt{}\Rm{}{\it{}idx}\nwendquote} as well.  Parameter {\Tt{}\Rm{}{\it{}metr}\nwendquote} defines the metric \(B(i,j)\) and
can be represented by a square {\Tt{}\Rm{}{\bf{}matrix}\nwendquote}, {\Tt{}\Rm{}{\bf{}tensormetric}\nwendquote} or
{\Tt{}\Rm{}{\bf{}indexed}\nwendquote} class object, optional parameter {\Tt{}\Rm{}{\it{}rl}\nwendquote} allows to
distinguish different Clifford algebras (which will commute each
other).  The last optional parameter {\Tt{}\Rm{}{\it{}anticommuting}\nwendquote} defines if the
anticommuting assumption (i.e.  \(e_i e_j + e_j e_i = 0\)) will be
used for contraction of Clifford units. If the {\Tt{}\Rm{}{\it{}metric}\nwendquote} is supplied
by a {\Tt{}\Rm{}{\bf{}matrix}\nwendquote} object, then the value of {\Tt{}\Rm{}{\it{}anticommuting}\nwendquote} is
calculated automatically and the supplied one will be ignored. One can
overcome this by giving {\Tt{}\Rm{}{\it{}metric}\nwendquote} through matrix wrapped into an
{\Tt{}\Rm{}{\bf{}indexed}\nwendquote} object.

Note that the call {\Tt{}\Rm{}{\it{}clifford\_unit}({\it{}mu}, {\it{}minkmetric}())\nwendquote} creates something very close to
{\Tt{}\Rm{}{\it{}dirac\_gamma}({\it{}mu})\nwendquote}, although {\Tt{}\Rm{}{\it{}dirac\_gamma}\nwendquote} have more
efficient simplification mechanism. The method {\Tt{}\Rm{}{\it{}clifford}::{\it{}get\_metric}()\nwendquote} returns
metric defining this Clifford number. The
method {\Tt{}\Rm{}{\it{}clifford}::{\it{}is\_anticommuting}()\nwendquote} returns the {\Tt{}\Rm{}{\it{}anticommuting}\nwendquote}
property of a unit.

If the matrix \(B(i, j)\) is in fact symmetric you may prefer to create
the Clifford algebra units with a call like that
\begin{webcode}{\Tt{}\Rm{}{\bf{}ex} {\it{}e} = {\it{}clifford\_unit}({\it{}mu}, {\bf{}indexed}({\it{}B}, {\it{}sy\_symm}(), {\it{}i}, {\it{}j}))\nwendquote};
\end{webcode}
since this may yield some further automatic simplifications. Again, for
a metric defined through a {\Tt{}\Rm{}{\bf{}matrix}\nwendquote} such a symmetry is detected
automatically.

Individual generators of a Clifford algebra can be accessed in several
ways. For example
\begin{webcode}{\Tt{}\Rm{}{\nwlbrace}\nwendquote}
    \ldots
    {\Tt{}\Rm{}{\bf{}varidx} {\it{}nu}({\bf{}symbol}({\tt{}"nu"}), 4)\nwendquote};
    {\Tt{}\Rm{}{\bf{}realsymbol} {\it{}s}({\tt{}"s"})\nwendquote};
    {\Tt{}\Rm{}{\bf{}ex} {\it{}M} = {\it{}diag\_matrix}({\bf{}lst}(1, -1, 0, {\it{}s}))\nwendquote};
    {\Tt{}\Rm{}{\bf{}ex} {\it{}e} = {\it{}clifford\_unit}({\it{}nu}, {\it{}M})\nwendquote};
    {\Tt{}\Rm{}{\bf{}ex} {\it{}e0} = {\it{}e}.{\it{}subs}({\it{}nu} \begin{math}\equiv\end{math} 0)\nwendquote};
    {\Tt{}\Rm{}{\bf{}ex} {\it{}e1} = {\it{}e}.{\it{}subs}({\it{}nu} \begin{math}\equiv\end{math} 1)\nwendquote};
    {\Tt{}\Rm{}{\bf{}ex} {\it{}e2} = {\it{}e}.{\it{}subs}({\it{}nu} \begin{math}\equiv\end{math} 2)\nwendquote};
    {\Tt{}\Rm{}{\bf{}ex} {\it{}e3} = {\it{}e}.{\it{}subs}({\it{}nu} \begin{math}\equiv\end{math} 3)\nwendquote};
    \ldots
  {\Tt{}\Rm{}{\nwrbrace}\nwendquote}
\end{webcode}
will produce four generators of a Clifford algebra with properties
\(e_0^2 = 1\),  \(e_1^2 = -1\), \(e_2^2 = 0\) and  \(e_3^2 = s\).

A similar effect can be achieved from the function
\begin{webcode}{\Tt{}\Rm{}{\bf{}ex} {\it{}lst\_to\_clifford}({\bf{}const} {\bf{}ex} & {\it{}v}, {\bf{}const} {\bf{}ex} & {\it{}mu}, {\bf{}const} {\bf{}ex} & {\it{}metr}, {\bf{}unsigned} {\bf{}char} {\it{}rl} = 0,\nwendquote}
      {\Tt{}\Rm{}{\bf{}bool} {\it{}anticommuting} = {\bf{}false});\nwendquote}
{\Tt{}\Rm{}{\bf{}ex} {\it{}lst\_to\_clifford}({\bf{}const} {\bf{}ex} & {\it{}v}, {\bf{}const} {\bf{}ex} & {\it{}e});\nwendquote}
\end{webcode}
which converts a list or vector {\Tt{}\Rm{}{\it{}v}\nwendquote}\(=(v_0, v_1, \ldots, v_n)\) into
the Clifford number \(v_0 e_0+v_1 e_1+\cdots + v_n e_n\) with \(e_k\)
directly supplied in the second form of the procedure. In the first
form the Clifford unit \(e_k\) is generated by {\Tt{}\Rm{}{\it{}clifford\_unit}({\it{}mu},\nwnewline
{\it{}metr}, {\it{}rl}, {\it{}anticommuting})\nwendquote}. The previous code may be rewritten with
help of {\Tt{}\Rm{}{\it{}lst\_to\_clifford}()\nwendquote} as follows
\begin{webcode}{\Tt{}\Rm{}{\nwlbrace}\nwendquote}
    \ldots
    {\Tt{}\Rm{}{\bf{}varidx} {\it{}nu}({\bf{}symbol}({\tt{}"nu"}), 4)\nwendquote};
    {\Tt{}\Rm{}{\bf{}realsymbol} {\it{}s}({\tt{}"s"})\nwendquote};
    {\Tt{}\Rm{}{\bf{}ex} {\it{}M} =  {\it{}diag\_matrix}({\bf{}lst}(1, -1, 0, {\it{}s}))\nwendquote};
    {\Tt{}\Rm{}{\bf{}ex} {\it{}e0} = {\it{}lst\_to\_clifford}({\bf{}lst}(1, 0, 0, 0), {\it{}nu}, {\it{}M})\nwendquote};
    {\Tt{}\Rm{}{\bf{}ex} {\it{}e1} = {\it{}lst\_to\_clifford}({\bf{}lst}(0, 1, 0, 0), {\it{}nu}, {\it{}M})\nwendquote};
    {\Tt{}\Rm{}{\bf{}ex} {\it{}e2} = {\it{}lst\_to\_clifford}({\bf{}lst}(0, 0, 1, 0), {\it{}nu}, {\it{}M})\nwendquote};
    {\Tt{}\Rm{}{\bf{}ex} {\it{}e3} = {\it{}lst\_to\_clifford}({\bf{}lst}(0, 0, 0, 1), {\it{}nu}, {\it{}M})\nwendquote};
  \ldots
  {\Tt{}\Rm{}{\nwrbrace}\nwendquote}
\end{webcode}
There is the inverse function
\begin{webcode}{\Tt{}\Rm{}{\bf{}lst} {\it{}clifford\_to\_lst}({\bf{}const} {\bf{}ex} & {\it{}e}, {\bf{}const} {\bf{}ex} & {\it{}c}, {\bf{}bool} {\it{}algebraic}={\bf{}true})\nwendquote};
\end{webcode}
which took an expression {\Tt{}\Rm{}{\it{}e}\nwendquote} and tries to find such a list
{\Tt{}\Rm{}{\it{}v}\nwendquote}\(=(v_0, v_1, \ldots, v_n)\) that \(e = v_0c_0 + v_1c_1 + \cdots
+ v_nc_n\) with respect to given Clifford units {\Tt{}\Rm{}{\it{}c}\nwendquote} and none of
\(v_k\) contains the Clifford units {\Tt{}\Rm{}{\it{}c}\nwendquote} (of course, this
may be impossible). This function can use an {\Tt{}\Rm{}{\it{}algebraic}\nwendquote} method
(default) or a symbolic one. In {\Tt{}\Rm{}{\it{}algebraic}\nwendquote} method \(v_k\) are calculated as
\((e c_k+ c_k e)/pow(c_k, 2)\).  If \(pow(c_k, 2)\) is zero or is not
{\Tt{}\Rm{}{\bf{}numeric}\nwendquote} for some \(k\)
then the method will be automatically changed to symbolic. The same effect
is obtained by the assignment ({\Tt{}\Rm{}{\it{}algebraic}={\bf{}false}\nwendquote}) in the procedure call.

There are several functions for (anti-)automorphisms of Clifford algebras:
\begin{webcode}{\Tt{}\Rm{}{\bf{}ex} {\it{}clifford\_prime}({\bf{}const} {\bf{}ex} & {\it{}e})\nwendquote}
{\Tt{}\Rm{}{\bf{}inline} {\bf{}ex} {\it{}clifford\_star}({\bf{}const} {\bf{}ex} & {\it{}e}) {\nwlbrace} {\bf{}return} {\it{}e}.{\it{}conjugate}(); {\nwrbrace}\nwendquote}
{\Tt{}\Rm{}{\bf{}inline} {\bf{}ex} {\it{}clifford\_bar}({\bf{}const} {\bf{}ex} & {\it{}e}) {\nwlbrace} {\bf{}return}   {\it{}clifford\_prime}({\it{}e}.{\it{}conjugate}()); {\nwrbrace}\nwendquote}
\end{webcode}
The automorphism of a Clifford algebra {\Tt{}\Rm{}{\it{}clifford\_prime}()\nwendquote} simply
changes signs of all Clifford units in the expression. The reversion
of a Clifford algebra {\Tt{}\Rm{}{\it{}clifford\_star}()\nwendquote} coincides with
{\Tt{}\Rm{}{\it{}conjugate}()\nwendquote} method and effectively reverses the order of Clifford
units in any product. Finally the main anti-automorphism
of a Clifford algebra {\Tt{}\Rm{}{\it{}clifford\_bar}()\nwendquote} is the composition of two
previous, i.e. makes the reversion and changes signs of all Clifford units
in a product. Names for this functions corresponds to notations
\(e'\), \(e^*\) and \(\bar{e}\) used in Clifford algebra
textbooks~\cite{BraDelSom82,Cnops02a,DelSomSou92}.

The function
\begin{webcode}{\Tt{}\Rm{}{\bf{}ex} {\it{}clifford\_norm}({\bf{}const} {\bf{}ex} & {\it{}e})\nwendquote};
\end{webcode}
calculates the  norm of Clifford number from the expression
\(\norm{e}^2=e\bar{e}\). The inverse of a Clifford expression is returned
by the function
\begin{webcode}{\Tt{}\Rm{}{\bf{}ex} {\it{}clifford\_inverse}({\bf{}const} {\bf{}ex} & {\it{}e})\nwendquote};
\end{webcode}
which calculates it as \(e^{-1}= e/\norm{e}^2\). If \(\norm{e}=0\) then an
exception is raised.

If a Clifford number happens to be a factor of
{\Tt{}\Rm{}{\it{}dirac\_ONE}()\nwendquote} then we can convert it to a ``real'' (non-Clifford)
expression by the function
\begin{webcode}{\Tt{}\Rm{}{\bf{}ex} {\it{}remove\_dirac\_ONE}({\bf{}const} {\bf{}ex} & {\it{}e})\nwendquote};
\end{webcode}
The function {\Tt{}\Rm{}{\it{}canonicalize\_clifford}()\nwendquote} works for a
generic Clifford algebra in a similar way as for Dirac gammas.

The last provided function is
\begin{webcode}{\Tt{}\Rm{}{\bf{}ex} {\it{}clifford\_moebius\_map}({\bf{}const} {\bf{}ex} & {\it{}a}, {\bf{}const} {\bf{}ex} & {\it{}b}, {\bf{}const} {\bf{}ex} & {\it{}c}, {\bf{}const} {\bf{}ex} & {\it{}d}, {\bf{}const} {\bf{}ex} & {\it{}v}, {\bf{}const} {\bf{}ex} & {\it{}G},\nwendquote}
                                 {\Tt{}\Rm{}{\bf{}unsigned} {\bf{}char} {\it{}rl} = 0, {\bf{}bool} {\it{}anticommuting} = {\bf{}false});\nwendquote}
{\Tt{}\Rm{}{\bf{}ex} {\it{}clifford\_moebius\_map}({\bf{}const} {\bf{}ex} & {\it{}M}, {\bf{}const} {\bf{}ex} & {\it{}v}, {\bf{}const} {\bf{}ex} & {\it{}G}, {\bf{}unsigned} {\bf{}char} {\it{}rl} = 0,\nwendquote}
                                 {\Tt{}\Rm{}{\bf{}bool} {\it{}anticommuting} = {\bf{}false});\nwendquote}
\end{webcode}
It takes a list or vector {\Tt{}\Rm{}{\it{}v}\nwendquote} and makes the M\"obius (conformal or
linear-fractional) transformation~\cite{Cnops02a}
\begin{displaymath}
  v\mapsto (av+b)(cv+d)^{-1} \quad \textrm{ defined by the matrix }
  M = \begin{pmatrix}
    a & b\\
    c & d
  \end{pmatrix}.
\end{displaymath}
The matrix may be given in two different forms---as one entity or by its four elements.
The last parameter {\Tt{}\Rm{}{\it{}G}\nwendquote} define the metric of the
surrounding (pseudo-)Euclidean space. This can be an indexed
object, tensormetric, matrix or a Clifford unit, in the later case the
optional parameters {\Tt{}\Rm{}{\it{}rl}\nwendquote} and {\Tt{}\Rm{}{\it{}anticommuting}\nwendquote} are ignored even if
supplied. The returned value of this
function is a list of components of the resulting vector.

Finally the function
\begin{webcode}{\Tt{}\Rm{}{\bf{}char} {\it{}clifford\_max\_label}({\bf{}const} {\bf{}ex} & {\it{}e}, {\bf{}bool} {\it{}ignore\_ONE} = {\bf{}false});\nwendquote}
\end{webcode}
can detect a presence of Clifford objects in the expression {\Tt{}\Rm{}{\it{}e}\nwendquote}: if
such objects are found it returns the maximal {\Tt{}\Rm{}{\it{}representation\_label}\nwendquote} of
them, otherwise {\Tt{}\Rm{}-1\nwendquote}. The optional parameter {\Tt{}\Rm{}{\it{}ignore\_ONE}\nwendquote} indicates if
{\Tt{}\Rm{}{\it{}dirac\_ONE}\nwendquote} objects should be ignored during the search.

\LaTeX\ output for Clifford units looks like \verb|\clifford[1]{e}^{{\nu}}}|,
where \texttt{1} is the {\Tt{}\Rm{}{\it{}representation\_label}\nwendquote} and \verb|\nu| is the
index of the corresponding unit. This provides a flexible typesetting
with a suitable defintion of the \verb|\clifford| command. For example, the
definition
\begin{verbatim}
    \newcommand{\clifford}[1][]{}
\end{verbatim}
typesets all Clifford units identically, while the alternative definition
\begin{verbatim}
    \newcommand{\clifford}[2][]{\ifcase #1 #2\or \tilde{#2} \or \breve{#2} \fi}
\end{verbatim}
prints units with {\Tt{}\Rm{}{\it{}representation\_label}=0\nwendquote} as
\(e\), with {\Tt{}\Rm{}{\it{}representation\_label}=1\nwendquote} as \(\tilde{e}\)
 and with {\Tt{}\Rm{}{\it{}representation\_label}=2\nwendquote} as
\(\breve{e}\).

\nwenddocs{}\nwbegindocs{10}\nwdocspar
\section{Main procedure}
Here is the main procedure, which has a very straightforward
structure.  This and next initialisation section is pretty standard.
The first usage of \GiNaC\ for Clifford algebras is in
Section~\ref{sec:cliff-algebra-calc}.
\nwenddocs{}\nwbegincode{11}\sublabel{NW3OF8UE-1p0Y9w-1}\nwmargintag{{\nwtagstyle{}\subpageref{NW3OF8UE-1p0Y9w-1}}}\moddef{*~{\nwtagstyle{}\subpageref{NW3OF8UE-1p0Y9w-1}}}\endmoddef\Rm{}\nwstartdeflinemarkup\nwprevnextdefs{\relax}{NW3OF8UE-1p0Y9w-2}\nwenddeflinemarkup
\LA{}Includes~{\nwtagstyle{}\subpageref{NW3OF8UE-ZKEBO-1}}\RA{}
\LA{}Definitions~{\nwtagstyle{}\subpageref{NW3OF8UE-4TccJC-1}}\RA{}
\LA{}Global items~{\nwtagstyle{}\subpageref{NW3OF8UE-4EE31V-1}}\RA{}
{\bf{}int} {\it{}main} ({\bf{}int} {\it{}argc}, {\bf{}char}\begin{math}\ast\end{math}\begin{math}\ast\end{math}{\it{}argv})\nwindexdefn{\nwixident{main}}{main}{NW3OF8UE-1p0Y9w-1}
{\nwlbrace}
    \LA{}C++ variables declaration~{\nwtagstyle{}\subpageref{NW3OF8UE-1cUGLU-1}}\RA{}
    \LA{}CiNaC variables declaration~{\nwtagstyle{}\subpageref{NW3OF8UE-2Cj0m9-1}}\RA{}
    \LA{}Pictures tuning~{\nwtagstyle{}\subpageref{NW3OF8UE-44zkoV-1}}\RA{}
    \LA{}Parabola parameters~{\nwtagstyle{}\subpageref{NW3OF8UE-2gOisp-1}}\RA{}

\nwalsodefined{\\{NW3OF8UE-1p0Y9w-2}\\{NW3OF8UE-1p0Y9w-3}\\{NW3OF8UE-1p0Y9w-4}}\nwnotused{*}\nwidentdefs{\\{{\nwixident{main}}{main}}}\nwendcode{}\nwbegindocs{12}Now we run a cycle over the three possible type of metric in two
dimensional space (i.e. {\Tt{}\Rm{}{\it{}elliptic}\nwendquote}, {\Tt{}\Rm{}{\it{}parabolic}\nwendquote}
and {\Tt{}\Rm{}{\it{}hyperbolic}\nwendquote}). For each space we initialise the corresponding
Clifford units, symbolically calculate various types of M\"obius
transforms as well as vector fields for three subgroups of \(\SL\).
\nwenddocs{}\nwbegincode{13}\sublabel{NW3OF8UE-1p0Y9w-2}\nwmargintag{{\nwtagstyle{}\subpageref{NW3OF8UE-1p0Y9w-2}}}\moddef{*~{\nwtagstyle{}\subpageref{NW3OF8UE-1p0Y9w-1}}}\plusendmoddef\Rm{}\nwstartdeflinemarkup\nwprevnextdefs{NW3OF8UE-1p0Y9w-1}{NW3OF8UE-1p0Y9w-3}\nwenddeflinemarkup
    {\bf{}for} ({\it{}metric} = {\it{}elliptic}; {\it{}metric} \begin{math}\leq\end{math} {\it{}hyperbolic}; {\it{}metric}\protect\PP) {\nwlbrace}
        {\it{}cout} \begin{math}\ll\end{math} {\it{}endl} \begin{math}\ll\end{math}{\it{}endl} \begin{math}\ll\end{math} {\tt{}"Metric is: "} \begin{math}\ll\end{math} {\it{}metric\_name}[{\it{}metric}] \begin{math}\ll\end{math} {\tt{}"."} \begin{math}\ll\end{math} {\it{}endl};
        \LA{}Initialise Clifford numbers~{\nwtagstyle{}\subpageref{NW3OF8UE-2xd863-1}}\RA{}
        \LA{}Calculation of Moebius transformations~{\nwtagstyle{}\subpageref{NW3OF8UE-1OFQOX-1}}\RA{}
        \LA{}Calculation of vector fields~{\nwtagstyle{}\subpageref{NW3OF8UE-4BpIv1-1}}\RA{}

\nwidentuses{\\{{\nwixident{elliptic}}{elliptic}}\\{{\nwixident{hyperbolic}}{hyperbolic}}\\{{\nwixident{metric}}{metric}}}\nwindexuse{\nwixident{elliptic}}{elliptic}{NW3OF8UE-1p0Y9w-2}\nwindexuse{\nwixident{hyperbolic}}{hyperbolic}{NW3OF8UE-1p0Y9w-2}\nwindexuse{\nwixident{metric}}{metric}{NW3OF8UE-1p0Y9w-2}\nwendcode{}\nwbegindocs{14}Then we run a cycle for three subgroups of \(\SL\) (i.e. \(A\),
\(N\), \(K\)). For all possible combinations of those with {\Tt{}\Rm{}{\it{}metric}\nwendquote} from
the surrounding cycle in the previous chunk we
\begin{enumerate}
\item build orbits of the subgroups and their transversal  curves;
\item two types of the Cayley transform images of all above curves;
\item check some formulae in the paper;
\end{enumerate}
We draw all pictures by substitution of numeric values into the
symbolic results obtained in the above chunks.
\nwenddocs{}\nwbegincode{15}\sublabel{NW3OF8UE-1p0Y9w-3}\nwmargintag{{\nwtagstyle{}\subpageref{NW3OF8UE-1p0Y9w-3}}}\moddef{*~{\nwtagstyle{}\subpageref{NW3OF8UE-1p0Y9w-1}}}\plusendmoddef\Rm{}\nwstartdeflinemarkup\nwprevnextdefs{NW3OF8UE-1p0Y9w-2}{NW3OF8UE-1p0Y9w-4}\nwenddeflinemarkup
        {\bf{}for} ({\it{}subgroup} = {\it{}subgroup\_A}; {\it{}subgroup} \begin{math}\leq\end{math} {\it{}subgroup\_K}; {\it{}subgroup}\protect\PP) {\nwlbrace}
        {\commopen} iteration over subgroups A, N and K {\commclose}
            \LA{}Drawing arrows~{\nwtagstyle{}\subpageref{NW3OF8UE-3M9GDT-1}}\RA{}
            \LA{}Building orbits~{\nwtagstyle{}\subpageref{NW3OF8UE-3Fz9v4-1}}\RA{}
            \LA{}Building transverses~{\nwtagstyle{}\subpageref{NW3OF8UE-38hDUf-1}}\RA{}
        {\nwrbrace}
    {\nwrbrace}

\nwidentuses{\\{{\nwixident{subgroup}}{subgroup}}\\{{\nwixident{subgroup{\_}A}}{subgroup:unA}}\\{{\nwixident{subgroup{\_}K}}{subgroup:unK}}}\nwindexuse{\nwixident{subgroup}}{subgroup}{NW3OF8UE-1p0Y9w-3}\nwindexuse{\nwixident{subgroup{\_}A}}{subgroup:unA}{NW3OF8UE-1p0Y9w-3}\nwindexuse{\nwixident{subgroup{\_}K}}{subgroup:unK}{NW3OF8UE-1p0Y9w-3}\nwendcode{}\nwbegindocs{16}Finally we draw eight frames which illustrates the
continuous transformation of the future part of the light cone into
the its past part~\cite[Figure~\ref{E-fig:future-to-past}]{Kisil05a}.
\nwenddocs{}\nwbegincode{17}\sublabel{NW3OF8UE-1p0Y9w-4}\nwmargintag{{\nwtagstyle{}\subpageref{NW3OF8UE-1p0Y9w-4}}}\moddef{*~{\nwtagstyle{}\subpageref{NW3OF8UE-1p0Y9w-1}}}\plusendmoddef\Rm{}\nwstartdeflinemarkup\nwprevnextdefs{NW3OF8UE-1p0Y9w-3}{\relax}\nwenddeflinemarkup
    \LA{}Build future-past transition~{\nwtagstyle{}\subpageref{NW3OF8UE-2KWnYy-1}}\RA{}
{\nwrbrace}

\nwendcode{}\nwbegindocs{18}\nwdocspar
\section{Auxiliary matter}
\label{sec:auxiliary-matter}

\nwenddocs{}\nwbegindocs{19}Some standard inclusions, but do not forget \GiNaC\ library!
\nwenddocs{}\nwbegincode{20}\sublabel{NW3OF8UE-ZKEBO-1}\nwmargintag{{\nwtagstyle{}\subpageref{NW3OF8UE-ZKEBO-1}}}\moddef{Includes~{\nwtagstyle{}\subpageref{NW3OF8UE-ZKEBO-1}}}\endmoddef\Rm{}\nwstartdeflinemarkup\nwusesondefline{\\{NW3OF8UE-1p0Y9w-1}}\nwenddeflinemarkup
{\bf{}\char35{}include}{\tt{} \begin{math}<\end{math}ginac/ginac.h\begin{math}>\end{math} // At least ver. 1.4.0!}
{\bf{}\char35{}include}{\tt{} \begin{math}<\end{math}cmath\begin{math}>\end{math}}

{\bf{}using} {\bf{}namespace} {\it{}std};
{\bf{}using} {\bf{}namespace} {\it{}GiNaC};

\nwused{\\{NW3OF8UE-1p0Y9w-1}}\nwendcode{}\nwbegindocs{21}\nwdocspar
\subsection{Defines}
\label{sec:defines}
 Some constants are defined here for a better readability of the code.
\nwenddocs{}\nwbegincode{22}\sublabel{NW3OF8UE-4TccJC-1}\nwmargintag{{\nwtagstyle{}\subpageref{NW3OF8UE-4TccJC-1}}}\moddef{Definitions~{\nwtagstyle{}\subpageref{NW3OF8UE-4TccJC-1}}}\endmoddef\Rm{}\nwstartdeflinemarkup\nwusesondefline{\\{NW3OF8UE-1p0Y9w-1}}\nwprevnextdefs{\relax}{NW3OF8UE-4TccJC-2}\nwenddeflinemarkup
// Defined constants
{\bf{}\char35{}define}{\tt{} elliptic 0}\nwindexdefn{\nwixident{elliptic}}{elliptic}{NW3OF8UE-4TccJC-1}
{\bf{}\char35{}define}{\tt{} parabolic 1}\nwindexdefn{\nwixident{parabolic}}{parabolic}{NW3OF8UE-4TccJC-1}
{\bf{}\char35{}define}{\tt{} hyperbolic 2}\nwindexdefn{\nwixident{hyperbolic}}{hyperbolic}{NW3OF8UE-4TccJC-1}
{\bf{}\char35{}define}{\tt{} subgroup\_A 0}\nwindexdefn{\nwixident{subgroup{\_}A}}{subgroup:unA}{NW3OF8UE-4TccJC-1}
{\bf{}\char35{}define}{\tt{} subgroup\_N 1}\nwindexdefn{\nwixident{subgroup{\_}N}}{subgroup:unN}{NW3OF8UE-4TccJC-1}
{\bf{}\char35{}define}{\tt{} subgroup\_K 2}\nwindexdefn{\nwixident{subgroup{\_}K}}{subgroup:unK}{NW3OF8UE-4TccJC-1}
{\bf{}\char35{}define}{\tt{} grey 0.6}\nwindexdefn{\nwixident{grey}}{grey}{NW3OF8UE-4TccJC-1}

\nwalsodefined{\\{NW3OF8UE-4TccJC-2}\\{NW3OF8UE-4TccJC-3}\\{NW3OF8UE-4TccJC-4}\\{NW3OF8UE-4TccJC-5}\\{NW3OF8UE-4TccJC-6}\\{NW3OF8UE-4TccJC-7}\\{NW3OF8UE-4TccJC-8}\\{NW3OF8UE-4TccJC-9}\\{NW3OF8UE-4TccJC-A}}\nwused{\\{NW3OF8UE-1p0Y9w-1}}\nwidentdefs{\\{{\nwixident{elliptic}}{elliptic}}\\{{\nwixident{grey}}{grey}}\\{{\nwixident{hyperbolic}}{hyperbolic}}\\{{\nwixident{parabolic}}{parabolic}}\\{{\nwixident{subgroup{\_}A}}{subgroup:unA}}\\{{\nwixident{subgroup{\_}K}}{subgroup:unK}}\\{{\nwixident{subgroup{\_}N}}{subgroup:unN}}}\nwendcode{}\nwbegindocs{23} Some macro definitions which we use to make more compact code.
They initialise variables, open and close curve description in the
\MetaPost\  file. Here is  initialisation of a new curve.
\nwenddocs{}\nwbegincode{24}\sublabel{NW3OF8UE-4TccJC-2}\nwmargintag{{\nwtagstyle{}\subpageref{NW3OF8UE-4TccJC-2}}}\moddef{Definitions~{\nwtagstyle{}\subpageref{NW3OF8UE-4TccJC-1}}}\plusendmoddef\Rm{}\nwstartdeflinemarkup\nwusesondefline{\\{NW3OF8UE-1p0Y9w-1}}\nwprevnextdefs{NW3OF8UE-4TccJC-1}{NW3OF8UE-4TccJC-3}\nwenddeflinemarkup
{\bf{}\char35{}define}{\tt{} init\_coord(X) upos[X] = 0;              \begin{math}\backslash\end{math}}\nwindexdefn{\nwixident{init{\_}coord}}{init:uncoord}{NW3OF8UE-4TccJC-2}
        {\it{}vpos}[{\it{}X}] = 0; \begin{math}\backslash\end{math}
        {\it{}udir}[{\it{}X}] = 1; \begin{math}\backslash\end{math}
        {\it{}vdir}[{\it{}X}] = 0; \begin{math}\backslash\end{math}
        {\it{}fprintf}({\it{}fileout}[{\it{}X}], {\tt{}"draw "})

\nwused{\\{NW3OF8UE-1p0Y9w-1}}\nwidentdefs{\\{{\nwixident{init{\_}coord}}{init:uncoord}}}\nwidentuses{\\{{\nwixident{fileout}}{fileout}}\\{{\nwixident{upos}}{upos}}}\nwindexuse{\nwixident{fileout}}{fileout}{NW3OF8UE-4TccJC-2}\nwindexuse{\nwixident{upos}}{upos}{NW3OF8UE-4TccJC-2}\nwendcode{}\nwbegindocs{25}This part is used to close a curve output in cases the end is
reached or curves passes the infinity.
\nwenddocs{}\nwbegincode{26}\sublabel{NW3OF8UE-4TccJC-3}\nwmargintag{{\nwtagstyle{}\subpageref{NW3OF8UE-4TccJC-3}}}\moddef{Definitions~{\nwtagstyle{}\subpageref{NW3OF8UE-4TccJC-1}}}\plusendmoddef\Rm{}\nwstartdeflinemarkup\nwusesondefline{\\{NW3OF8UE-1p0Y9w-1}}\nwprevnextdefs{NW3OF8UE-4TccJC-2}{NW3OF8UE-4TccJC-4}\nwenddeflinemarkup
{\bf{}\char35{}define}{\tt{} close\_curve(X) fprintf(fileout[X], "(a
    {\it{}upos}[{\it{}X}], {\it{}vpos}[{\it{}X}], {\it{}color\_grade}, {\it{}color\_name}[{\it{}subgroup}])
{\bf{}\char35{}define}{\tt{} put\_draw(X) fprintf(fileout[X], "\begin{math}\backslash\end{math}ndraw ")}\nwindexdefn{\nwixident{put{\_}draw}}{put:undraw}{NW3OF8UE-4TccJC-3}

\nwused{\\{NW3OF8UE-1p0Y9w-1}}\nwidentdefs{\\{{\nwixident{close{\_}curve}}{close:uncurve}}\\{{\nwixident{put{\_}draw}}{put:undraw}}}\nwidentuses{\\{{\nwixident{color{\_}name}}{color:unname}}\\{{\nwixident{fileout}}{fileout}}\\{{\nwixident{subgroup}}{subgroup}}\\{{\nwixident{u}}{u}}\\{{\nwixident{upos}}{upos}}}\nwindexuse{\nwixident{color{\_}name}}{color:unname}{NW3OF8UE-4TccJC-3}\nwindexuse{\nwixident{fileout}}{fileout}{NW3OF8UE-4TccJC-3}\nwindexuse{\nwixident{subgroup}}{subgroup}{NW3OF8UE-4TccJC-3}\nwindexuse{\nwixident{u}}{u}{NW3OF8UE-4TccJC-3}\nwindexuse{\nwixident{upos}}{upos}{NW3OF8UE-4TccJC-3}\nwendcode{}\nwbegindocs{27}Here is the common part of code which is used for outputs a
segment  to files.
\nwenddocs{}\nwbegincode{28}\sublabel{NW3OF8UE-4TccJC-4}\nwmargintag{{\nwtagstyle{}\subpageref{NW3OF8UE-4TccJC-4}}}\moddef{Definitions~{\nwtagstyle{}\subpageref{NW3OF8UE-4TccJC-1}}}\plusendmoddef\Rm{}\nwstartdeflinemarkup\nwusesondefline{\\{NW3OF8UE-1p0Y9w-1}}\nwprevnextdefs{NW3OF8UE-4TccJC-3}{NW3OF8UE-4TccJC-5}\nwenddeflinemarkup
{\bf{}\char35{}define}{\tt{} put\_point(X) if (inversion) \begin{math}\backslash\end{math}}\nwindexdefn{\nwixident{put{\_}point}}{put:unpoint}{NW3OF8UE-4TccJC-4}
     {\it{}fprintf}({\it{}fileout}[{\it{}X}], {\tt{}"(a
 {\bf{}else} {\bf{}if} ({\it{}direct})                                                       \begin{math}\backslash\end{math}
     {\it{}fprintf}({\it{}fileout}[{\it{}X}], {\tt{}"(a
 {\bf{}else}                                                                   \begin{math}\backslash\end{math}
     {\it{}fprintf}({\it{}fileout}[{\it{}X}], {\tt{}"(a
{\bf{}\char35{}define}{\tt{} renew\_curve(Y)       close\_curve(Y);    \begin{math}\backslash\end{math}}\nwindexdefn{\nwixident{renew{\_}curve}}{renew:uncurve}{NW3OF8UE-4TccJC-4}
 {\it{}put\_draw}({\it{}Y})

\nwused{\\{NW3OF8UE-1p0Y9w-1}}\nwidentdefs{\\{{\nwixident{put{\_}point}}{put:unpoint}}\\{{\nwixident{renew{\_}curve}}{renew:uncurve}}}\nwidentuses{\\{{\nwixident{close{\_}curve}}{close:uncurve}}\\{{\nwixident{fileout}}{fileout}}\\{{\nwixident{put{\_}draw}}{put:undraw}}\\{{\nwixident{u}}{u}}\\{{\nwixident{upos}}{upos}}}\nwindexuse{\nwixident{close{\_}curve}}{close:uncurve}{NW3OF8UE-4TccJC-4}\nwindexuse{\nwixident{fileout}}{fileout}{NW3OF8UE-4TccJC-4}\nwindexuse{\nwixident{put{\_}draw}}{put:undraw}{NW3OF8UE-4TccJC-4}\nwindexuse{\nwixident{u}}{u}{NW3OF8UE-4TccJC-4}\nwindexuse{\nwixident{upos}}{upos}{NW3OF8UE-4TccJC-4}\nwendcode{}\nwbegindocs{29}We should make a rough check that the curve is still in the bounded
area, if it cross infinity then such line should be discontinued and
started from a new. Our bound are few times bigger that the real
picture, the excellent cutting within the desired limits is done by
\MetaPost\ itself with the \texttt{clip currentpicture to \ldots;}
command.

Besides some outer margins we put different types of bound depending
from the nature of objects: sometimes it is limited to the upper half
plane, sometimes to hyperbolic unit disk. The necessity of such checks
in the hyperbolic case is explained
in~\cite[\S~\ref{E-sec:invar-upper-half}]{Kisil05a}.
\nwenddocs{}\nwbegincode{30}\sublabel{NW3OF8UE-4TccJC-5}\nwmargintag{{\nwtagstyle{}\subpageref{NW3OF8UE-4TccJC-5}}}\moddef{Definitions~{\nwtagstyle{}\subpageref{NW3OF8UE-4TccJC-1}}}\plusendmoddef\Rm{}\nwstartdeflinemarkup\nwusesondefline{\\{NW3OF8UE-1p0Y9w-1}}\nwprevnextdefs{NW3OF8UE-4TccJC-4}{NW3OF8UE-4TccJC-6}\nwenddeflinemarkup
{\bf{}\char35{}define}{\tt{} if\_in\_limits(X) if ( (abs(u\_res.to\_double()) \begin{math}<\end{math}= ulim) \begin{math}\backslash\end{math}}\nwindexdefn{\nwixident{if{\_}in{\_}limits}}{if:unin:unlimits}{NW3OF8UE-4TccJC-5}
    \begin{math}\wedge\end{math} ({\it{}abs}({\it{}v\_res}.{\it{}to\_double}()) \begin{math}\leq\end{math} {\it{}vlim}) \begin{math}\backslash\end{math}
    \begin{math}\wedge\end{math} (({\it{}metric} \begin{math}\neq\end{math} {\it{}hyperbolic}) \begin{math}\backslash\end{math}
        \begin{math}\vee\end{math} {\it{}inversion}                                                    \begin{math}\backslash\end{math}
        \begin{math}\vee\end{math} ( \begin{math}\neg\end{math}{\it{}cayley} \begin{math}\wedge\end{math} ({\it{}v\_res}.{\it{}is\_positive}() \begin{math}\vee\end{math} {\it{}v\_res}.{\it{}is\_zero}()))       \begin{math}\backslash\end{math}
        \begin{math}\vee\end{math} ( {\it{}cayley} \begin{math}\wedge\end{math} \begin{math}\neg\end{math}{\it{}ex\_to}\begin{math}<\end{math}{\bf{}numeric}\begin{math}>\end{math}(-{\it{}pow}({\it{}u\_res},2)+{\it{}pow}({\it{}v\_res},2)-1.001).{\it{}is\_positive}()))) {\nwlbrace} \begin{math}\backslash\end{math}
                {\it{}upos}[{\it{}X}] = {\it{}u\_res}.{\it{}to\_double}();                            \begin{math}\backslash\end{math}
                {\it{}vpos}[{\it{}X}] = {\it{}v\_res}.{\it{}to\_double}();                \begin{math}\backslash\end{math}
                {\bf{}ex} {\it{}Vect} = {\it{}dV}[{\it{}subgroup}][{\it{}X}].{\it{}subs}({\bf{}lst}({\it{}x} \begin{math}\equiv\end{math} {\it{}u\_res}, {\it{}y} \begin{math}\equiv\end{math} {\it{}v\_res})); \begin{math}\backslash\end{math}
                {\it{}udir}[{\it{}X}] = {\it{}ex\_to}\begin{math}<\end{math}{\bf{}numeric}\begin{math}>\end{math}({\it{}Vect}.{\it{}op}(0)).{\it{}to\_double}();  \begin{math}\backslash\end{math}
                {\it{}vdir}[{\it{}X}] = {\it{}ex\_to}\begin{math}<\end{math}{\bf{}numeric}\begin{math}>\end{math}({\it{}Vect}.{\it{}op}(1)).{\it{}to\_double}();  \begin{math}\backslash\end{math}
                {\it{}put\_point}({\it{}X});                                           \begin{math}\backslash\end{math}
        {\nwrbrace} {\bf{}else} {\nwlbrace}                                                        \begin{math}\backslash\end{math}
                {\it{}renew\_curve}({\it{}X});                                         \begin{math}\backslash\end{math}
        {\nwrbrace}

\nwused{\\{NW3OF8UE-1p0Y9w-1}}\nwidentdefs{\\{{\nwixident{if{\_}in{\_}limits}}{if:unin:unlimits}}}\nwidentuses{\\{{\nwixident{hyperbolic}}{hyperbolic}}\\{{\nwixident{metric}}{metric}}\\{{\nwixident{numeric}}{numeric}}\\{{\nwixident{put{\_}point}}{put:unpoint}}\\{{\nwixident{renew{\_}curve}}{renew:uncurve}}\\{{\nwixident{subgroup}}{subgroup}}\\{{\nwixident{ulim}}{ulim}}\\{{\nwixident{upos}}{upos}}}\nwindexuse{\nwixident{hyperbolic}}{hyperbolic}{NW3OF8UE-4TccJC-5}\nwindexuse{\nwixident{metric}}{metric}{NW3OF8UE-4TccJC-5}\nwindexuse{\nwixident{numeric}}{numeric}{NW3OF8UE-4TccJC-5}\nwindexuse{\nwixident{put{\_}point}}{put:unpoint}{NW3OF8UE-4TccJC-5}\nwindexuse{\nwixident{renew{\_}curve}}{renew:uncurve}{NW3OF8UE-4TccJC-5}\nwindexuse{\nwixident{subgroup}}{subgroup}{NW3OF8UE-4TccJC-5}\nwindexuse{\nwixident{ulim}}{ulim}{NW3OF8UE-4TccJC-5}\nwindexuse{\nwixident{upos}}{upos}{NW3OF8UE-4TccJC-5}\nwendcode{}\nwbegindocs{31}Then a curve is going through infinity we catch the exception, close
the corresponding {\Tt{}\Rm{}{\it{}draw}\nwendquote} statement of \MetaPost\ and start a new one
from the next point.
\nwenddocs{}\nwbegincode{32}\sublabel{NW3OF8UE-4TccJC-6}\nwmargintag{{\nwtagstyle{}\subpageref{NW3OF8UE-4TccJC-6}}}\moddef{Definitions~{\nwtagstyle{}\subpageref{NW3OF8UE-4TccJC-1}}}\plusendmoddef\Rm{}\nwstartdeflinemarkup\nwusesondefline{\\{NW3OF8UE-1p0Y9w-1}}\nwprevnextdefs{NW3OF8UE-4TccJC-5}{NW3OF8UE-4TccJC-7}\nwenddeflinemarkup
{\bf{}\char35{}define}{\tt{} catch\_handle(X) cerr \begin{math}<\end{math}\begin{math}<\end{math} "*** Got problem: " \begin{math}<\end{math}\begin{math}<\end{math} p.what() \begin{math}<\end{math}\begin{math}<\end{math} endl; \begin{math}\backslash\end{math}}\nwindexdefn{\nwixident{catch{\_}handle}}{catch:unhandle}{NW3OF8UE-4TccJC-6}
        {\it{}renew\_curve}({\it{}X})
\nwused{\\{NW3OF8UE-1p0Y9w-1}}\nwidentdefs{\\{{\nwixident{catch{\_}handle}}{catch:unhandle}}}\nwidentuses{\\{{\nwixident{renew{\_}curve}}{renew:uncurve}}}\nwindexuse{\nwixident{renew{\_}curve}}{renew:uncurve}{NW3OF8UE-4TccJC-6}\nwendcode{}\nwbegindocs{33}Extracting of numerical values out of Moebius transformations
\nwenddocs{}\nwbegincode{34}\sublabel{NW3OF8UE-4TccJC-7}\nwmargintag{{\nwtagstyle{}\subpageref{NW3OF8UE-4TccJC-7}}}\moddef{Definitions~{\nwtagstyle{}\subpageref{NW3OF8UE-4TccJC-1}}}\plusendmoddef\Rm{}\nwstartdeflinemarkup\nwusesondefline{\\{NW3OF8UE-1p0Y9w-1}}\nwprevnextdefs{NW3OF8UE-4TccJC-6}{NW3OF8UE-4TccJC-8}\nwenddeflinemarkup
{\bf{}\char35{}define}{\tt{} get\_components u\_res = ex\_to\begin{math}<\end{math}numeric\begin{math}>\end{math}(res.op(0).evalf());   \begin{math}\backslash\end{math}}\nwindexdefn{\nwixident{get{\_}components}}{get:uncomponents}{NW3OF8UE-4TccJC-7}
                                                                   {\it{}v\_res} = {\it{}ex\_to}\begin{math}<\end{math}{\bf{}numeric}\begin{math}>\end{math}({\it{}res}.{\it{}op}(1).{\it{}evalf}())

\nwused{\\{NW3OF8UE-1p0Y9w-1}}\nwidentdefs{\\{{\nwixident{get{\_}components}}{get:uncomponents}}}\nwidentuses{\\{{\nwixident{numeric}}{numeric}}}\nwindexuse{\nwixident{numeric}}{numeric}{NW3OF8UE-4TccJC-7}\nwendcode{}\nwbegindocs{35}To make an accurate drawing we calculate the direction of a transverse
line out of symbolic vector fields calculation done before.
\nwenddocs{}\nwbegincode{36}\sublabel{NW3OF8UE-4TccJC-8}\nwmargintag{{\nwtagstyle{}\subpageref{NW3OF8UE-4TccJC-8}}}\moddef{Definitions~{\nwtagstyle{}\subpageref{NW3OF8UE-4TccJC-1}}}\plusendmoddef\Rm{}\nwstartdeflinemarkup\nwusesondefline{\\{NW3OF8UE-1p0Y9w-1}}\nwprevnextdefs{NW3OF8UE-4TccJC-7}{NW3OF8UE-4TccJC-9}\nwenddeflinemarkup
{\bf{}\char35{}define}{\tt{} transverse\_dir(X)   if (!direct) {\nwlbrace} \begin{math}\backslash\end{math}}\nwindexdefn{\nwixident{transverse{\_}dir}}{transverse:undir}{NW3OF8UE-4TccJC-8}
        {\it{}trans\_uf} = {\it{}ex\_to}\begin{math}<\end{math}{\bf{}numeric}\begin{math}>\end{math}({\it{}trans\_dir\_sub}[{\it{}X}].{\it{}op}(0).{\it{}subs}({\it{}a\_node}).{\it{}evalf}()).{\it{}to\_double}(); \begin{math}\backslash\end{math}
        {\it{}trans\_vf} = {\it{}ex\_to}\begin{math}<\end{math}{\bf{}numeric}\begin{math}>\end{math}({\it{}trans\_dir\_sub}[{\it{}X}].{\it{}op}(1).{\it{}subs}({\it{}a\_node}).{\it{}evalf}()).{\it{}to\_double}(); \begin{math}\backslash\end{math}
        {\bf{}if} ({\it{}trans\_uf} \begin{math}\equiv\end{math} {\it{}INFINITY}) {\nwlbrace} {\it{}trans\_uf} = 1; {\it{}trans\_vf} = 0; {\nwrbrace} \begin{math}\backslash\end{math}
        {\bf{}else} {\bf{}if} ({\it{}trans\_uf} \begin{math}\equiv\end{math} -{\it{}INFINITY}) {\nwlbrace} {\it{}trans\_uf} = -1; {\it{}trans\_vf} = 0; {\nwrbrace} \begin{math}\backslash\end{math}
        {\bf{}else} {\bf{}if} ({\it{}trans\_vf} \begin{math}\equiv\end{math} {\it{}INFINITY}) {\nwlbrace} {\it{}trans\_uf} =0; {\it{}trans\_vf} = 1; {\nwrbrace} \begin{math}\backslash\end{math}
        {\bf{}else} {\bf{}if} ({\it{}trans\_vf} \begin{math}\equiv\end{math} -{\it{}INFINITY}) {\nwlbrace} {\it{}trans\_uf} =0; {\it{}trans\_vf} = -1; {\nwrbrace} \begin{math}\backslash\end{math}
        {\bf{}else} {\bf{}if} ({\it{}abs}({\it{}trans\_uf})+{\it{}abs}({\it{}trans\_vf}) \begin{math}>\end{math} 100) {\nwlbrace} \begin{math}\backslash\end{math}
          {\bf{}double} {\it{}r} = {\it{}sqrt}({\it{}trans\_uf} \begin{math}\ast\end{math} {\it{}trans\_uf} + {\it{}trans\_vf} \begin{math}\ast\end{math} {\it{}trans\_vf}); \begin{math}\backslash\end{math}
          {\it{}trans\_uf} \begin{math}\div\end{math}= {\it{}r}; {\it{}trans\_vf} \begin{math}\div\end{math}= {\it{}r}; \begin{math}\backslash\end{math}
        {\nwrbrace} \begin{math}\backslash\end{math}
{\nwrbrace}

\nwused{\\{NW3OF8UE-1p0Y9w-1}}\nwidentdefs{\\{{\nwixident{transverse{\_}dir}}{transverse:undir}}}\nwidentuses{\\{{\nwixident{numeric}}{numeric}}}\nwindexuse{\nwixident{numeric}}{numeric}{NW3OF8UE-4TccJC-8}\nwendcode{}\nwbegindocs{37}Some global variables which is convenient to use for the
{\Tt{}\Rm{}{\it{}openfile}()\nwendquote} function below.
\nwenddocs{}\nwbegincode{38}\sublabel{NW3OF8UE-4EE31V-1}\nwmargintag{{\nwtagstyle{}\subpageref{NW3OF8UE-4EE31V-1}}}\moddef{Global items~{\nwtagstyle{}\subpageref{NW3OF8UE-4EE31V-1}}}\endmoddef\Rm{}\nwstartdeflinemarkup\nwusesondefline{\\{NW3OF8UE-1p0Y9w-1}}\nwprevnextdefs{\relax}{NW3OF8UE-4EE31V-2}\nwenddeflinemarkup
{\bf{}int}  {\it{}subgroup}, {\it{}metric}; // Subgroup iterator and sign of the metric of the space\nwindexdefn{\nwixident{subgroup}}{subgroup}{NW3OF8UE-4EE31V-1}\nwindexdefn{\nwixident{metric}}{metric}{NW3OF8UE-4EE31V-1}
{\bf{}numeric} {\it{}signum};
{\bf{}char} {\it{}sgroup}[]={\tt{}"ANK"}, {\it{}metric\_name}[]={\tt{}"eph"}; // Names used for a readable output\nwindexdefn{\nwixident{sgroup}}{sgroup}{NW3OF8UE-4EE31V-1}

\nwalsodefined{\\{NW3OF8UE-4EE31V-2}}\nwused{\\{NW3OF8UE-1p0Y9w-1}}\nwidentdefs{\\{{\nwixident{metric}}{metric}}\\{{\nwixident{sgroup}}{sgroup}}\\{{\nwixident{subgroup}}{subgroup}}}\nwidentuses{\\{{\nwixident{numeric}}{numeric}}}\nwindexuse{\nwixident{numeric}}{numeric}{NW3OF8UE-4EE31V-1}\nwendcode{}\nwbegindocs{39}Procedure
{\Tt{}\Rm{}{\it{}openfile}()\nwendquote} is used for opening numerous data files with predefined
template of name.
\nwenddocs{}\nwbegincode{40}\sublabel{NW3OF8UE-4EE31V-2}\nwmargintag{{\nwtagstyle{}\subpageref{NW3OF8UE-4EE31V-2}}}\moddef{Global items~{\nwtagstyle{}\subpageref{NW3OF8UE-4EE31V-1}}}\plusendmoddef\Rm{}\nwstartdeflinemarkup\nwusesondefline{\\{NW3OF8UE-1p0Y9w-1}}\nwprevnextdefs{NW3OF8UE-4EE31V-1}{\relax}\nwenddeflinemarkup
{\it{}FILE} \begin{math}\ast\end{math}{\it{}openfile}({\bf{}const} {\bf{}char} \begin{math}\ast\end{math}{\it{}F}) {\nwlbrace}\nwindexdefn{\nwixident{openfile}}{openfile}{NW3OF8UE-4EE31V-2}
    {\bf{}char}  {\it{}filename}[]={\tt{}"cayley-t-k-e.d"}, {\it{}templ}[]={\tt{}"cayley-t-
    {\bf{}char} \begin{math}\ast\end{math}{\it{}Sfilename}={\it{}filename}, \begin{math}\ast\end{math}{\it{}Stempl}={\it{}templ};
    {\it{}strcat}({\it{}strcpy}({\it{}Stempl}, {\it{}F}), {\tt{}"-
    {\it{}sprintf}({\it{}Sfilename}, {\it{}Stempl}, &{\it{}sgroup}[{\it{}subgroup}], &{\it{}metric\_name}[{\it{}metric}]);
    {\bf{}return} {\it{}fopen}({\it{}Sfilename}, {\tt{}"w"});
{\nwrbrace}

\nwused{\\{NW3OF8UE-1p0Y9w-1}}\nwidentdefs{\\{{\nwixident{openfile}}{openfile}}}\nwidentuses{\\{{\nwixident{metric}}{metric}}\\{{\nwixident{sgroup}}{sgroup}}\\{{\nwixident{subgroup}}{subgroup}}}\nwindexuse{\nwixident{metric}}{metric}{NW3OF8UE-4EE31V-2}\nwindexuse{\nwixident{sgroup}}{sgroup}{NW3OF8UE-4EE31V-2}\nwindexuse{\nwixident{subgroup}}{subgroup}{NW3OF8UE-4EE31V-2}\nwendcode{}\nwbegindocs{41}\nwdocspar
\subsection{Variables}
\label{sec:variables}

First we define  variables from the standard \CPP\ classes.
\nwenddocs{}\nwbegincode{42}\sublabel{NW3OF8UE-1cUGLU-1}\nwmargintag{{\nwtagstyle{}\subpageref{NW3OF8UE-1cUGLU-1}}}\moddef{C++ variables declaration~{\nwtagstyle{}\subpageref{NW3OF8UE-1cUGLU-1}}}\endmoddef\Rm{}\nwstartdeflinemarkup\nwusesondefline{\\{NW3OF8UE-1p0Y9w-1}}\nwenddeflinemarkup
{\it{}FILE}  \begin{math}\ast\end{math}{\it{}fileout}[3]; {\commopen} files to pass results to \begin{math}\backslash\end{math}MetaPost\begin{math}\backslash\end{math}  {\commclose}\nwindexdefn{\nwixident{fileout}}{fileout}{NW3OF8UE-1cUGLU-1}
{\bf{}static} {\bf{}char} \begin{math}\ast\end{math}{\it{}color\_name}[]={\nwlbrace} {\tt{}"hyp"}, {\tt{}"par"}, {\tt{}"ell"}, {\tt{}"white"}{\nwrbrace}, // \(A\), \(N\), \(K\) subgroup colours\nwindexdefn{\nwixident{color{\_}name}}{color:unname}{NW3OF8UE-1cUGLU-1}
\begin{math}\ast\end{math}{\it{}formula}[]={\nwlbrace}{\tt{}"{\char92}nDistance to center is:"}, {\tt{}"{\char92}nDirectrice is:"},
            {\tt{}"{\char92}nDifference to foci is:"}{\nwrbrace};
{\bf{}bool} {\it{}direct} = {\bf{}true}, // Is it orbit or transverse?
    {\it{}cayley} = {\bf{}false}, // Is it the Cayley transform image?
    {\it{}inversion} = {\bf{}false}; //Is it future-past inversion?
{\bf{}double} {\it{}u}, {\it{}v}, {\it{}upos}[3], {\it{}vpos}[3], {\it{}udir}[3], {\it{}vdir}[3], {\it{}vval} = 0, // coordinates of point, vector, etc\nwindexdefn{\nwixident{u}}{u}{NW3OF8UE-1cUGLU-1}\nwindexdefn{\nwixident{v}}{v}{NW3OF8UE-1cUGLU-1}\nwindexdefn{\nwixident{upos}}{upos}{NW3OF8UE-1cUGLU-1}
    {\it{}color\_grade}, {\it{}focal\_f}[2] = {\nwlbrace}0, 0{\nwrbrace},
    {\it{}trans\_uf}=1, {\it{}trans\_vf}=0;

\nwused{\\{NW3OF8UE-1p0Y9w-1}}\nwidentdefs{\\{{\nwixident{color{\_}name}}{color:unname}}\\{{\nwixident{fileout}}{fileout}}\\{{\nwixident{u}}{u}}\\{{\nwixident{upos}}{upos}}\\{{\nwixident{v}}{v}}}\nwidentuses{\\{{\nwixident{subgroup}}{subgroup}}}\nwindexuse{\nwixident{subgroup}}{subgroup}{NW3OF8UE-1cUGLU-1}\nwendcode{}\nwbegindocs{43}Then other variables of \GiNaC\ types are defined as well. They
are needed for numeric and symbolic calculations
\nwenddocs{}\nwbegincode{44}\sublabel{NW3OF8UE-2Cj0m9-1}\nwmargintag{{\nwtagstyle{}\subpageref{NW3OF8UE-2Cj0m9-1}}}\moddef{CiNaC variables declaration~{\nwtagstyle{}\subpageref{NW3OF8UE-2Cj0m9-1}}}\endmoddef\Rm{}\nwstartdeflinemarkup\nwusesondefline{\\{NW3OF8UE-1p0Y9w-1}}\nwprevnextdefs{\relax}{NW3OF8UE-2Cj0m9-2}\nwenddeflinemarkup
{\bf{}varidx}  {\it{}nu}({\bf{}symbol}({\tt{}"nu"}, {\tt{}"{\char92}{\char92}nu"}), 2), {\it{}mu}({\bf{}symbol}({\tt{}"mu"}, {\tt{}"{\char92}{\char92}mu"}), 2),
    {\it{}psi}({\bf{}symbol}({\tt{}"psi"}, {\tt{}"{\char92}{\char92}psi"}),2),  {\it{}xi}({\bf{}symbol}({\tt{}"xi"}, {\tt{}"{\char92}{\char92}xi"}), 2);
{\bf{}realsymbol} {\it{}x}({\tt{}"x"}), {\it{}y}({\tt{}"y"}), {\it{}t}({\tt{}"t"}), // for symbolic calculations
    {\it{}a}({\tt{}"a"}), {\it{}b}({\tt{}"b"}), {\it{}c}({\tt{}"c"}), // parameters of the parabola \(v = au\sp2 + bu + c\)
    {\it{}tr\_u}({\tt{}"U"}), {\it{}tr\_v}({\tt{}"V"}); // Vector ot the tranverse direction
{\bf{}lst}  {\it{}a\_node}, {\it{}soln}[2], {\it{}a\_trans};
\nwalsodefined{\\{NW3OF8UE-2Cj0m9-2}\\{NW3OF8UE-2Cj0m9-3}\\{NW3OF8UE-2Cj0m9-4}}\nwused{\\{NW3OF8UE-1p0Y9w-1}}\nwidentuses{\\{{\nwixident{v}}{v}}}\nwindexuse{\nwixident{v}}{v}{NW3OF8UE-2Cj0m9-1}\nwendcode{}\nwbegindocs{45}\nwdocspar
\nwenddocs{}\nwbegincode{46}\sublabel{NW3OF8UE-2Cj0m9-2}\nwmargintag{{\nwtagstyle{}\subpageref{NW3OF8UE-2Cj0m9-2}}}\moddef{CiNaC variables declaration~{\nwtagstyle{}\subpageref{NW3OF8UE-2Cj0m9-1}}}\plusendmoddef\Rm{}\nwstartdeflinemarkup\nwusesondefline{\\{NW3OF8UE-1p0Y9w-1}}\nwprevnextdefs{NW3OF8UE-2Cj0m9-1}{NW3OF8UE-2Cj0m9-3}\nwenddeflinemarkup
{\bf{}matrix} {\it{}M}(2, 2),         // The metric of the vector space
    {\it{}C}(2, 2), {\it{}C1}(2, 2), {\it{}CI}(2, 2), {\it{}C1I}(2, 2), // Two versions of the Cayley transform
    {\it{}T}(2, 2), {\it{}TI}(2, 2),          // The map from first to second Cayley transform
    {\it{}Jacob}[3][3]={\nwlbrace}{\bf{}matrix}(2,2){\nwrbrace}, // Jacobian of the Moebius transformation
    {\it{}trans\_dir}[3][3]={\nwlbrace}{\bf{}matrix}(2,1){\nwrbrace}, // Components of transverse direction
    {\it{}trans\_dir\_sub}[3]={\nwlbrace}{\bf{}matrix}(2,1){\nwrbrace}; // Components of transverse direction substituted by a subgroup
\nwused{\\{NW3OF8UE-1p0Y9w-1}}\nwidentuses{\\{{\nwixident{metric}}{metric}}\\{{\nwixident{subgroup}}{subgroup}}}\nwindexuse{\nwixident{metric}}{metric}{NW3OF8UE-2Cj0m9-2}\nwindexuse{\nwixident{subgroup}}{subgroup}{NW3OF8UE-2Cj0m9-2}\nwendcode{}\nwbegindocs{47}\nwdocspar
\nwenddocs{}\nwbegincode{48}\sublabel{NW3OF8UE-2Cj0m9-3}\nwmargintag{{\nwtagstyle{}\subpageref{NW3OF8UE-2Cj0m9-3}}}\moddef{CiNaC variables declaration~{\nwtagstyle{}\subpageref{NW3OF8UE-2Cj0m9-1}}}\plusendmoddef\Rm{}\nwstartdeflinemarkup\nwusesondefline{\\{NW3OF8UE-1p0Y9w-1}}\nwprevnextdefs{NW3OF8UE-2Cj0m9-2}{NW3OF8UE-2Cj0m9-4}\nwenddeflinemarkup
{\bf{}numeric} {\it{}u\_res}, {\it{}v\_res}, //coordinates of the Moebius transform
    {\it{}up}[3][2] = {\nwlbrace}0, 0, 0, 0, 0, 0{\nwrbrace}, // saved values of coordinates of the parabola
    {\it{}vp}[3][2] = {\nwlbrace}0, 0, 0, 0, 0, 0{\nwrbrace};
{\bf{}const} {\bf{}numeric} {\it{}half}(1, 2);\nwindexdefn{\nwixident{numeric}}{numeric}{NW3OF8UE-2Cj0m9-3}
\nwused{\\{NW3OF8UE-1p0Y9w-1}}\nwidentdefs{\\{{\nwixident{numeric}}{numeric}}}\nwendcode{}\nwbegindocs{49}\nwdocspar
\nwenddocs{}\nwbegincode{50}\sublabel{NW3OF8UE-2Cj0m9-4}\nwmargintag{{\nwtagstyle{}\subpageref{NW3OF8UE-2Cj0m9-4}}}\moddef{CiNaC variables declaration~{\nwtagstyle{}\subpageref{NW3OF8UE-2Cj0m9-1}}}\plusendmoddef\Rm{}\nwstartdeflinemarkup\nwusesondefline{\\{NW3OF8UE-1p0Y9w-1}}\nwprevnextdefs{NW3OF8UE-2Cj0m9-3}{\relax}\nwenddeflinemarkup
{\bf{}ex} {\it{}res}, {\it{}e}, {\it{}e0}, {\it{}e1},
    {\it{}Moebius}[3][5], {\it{}dV}[3][3],
// indexed by [{\it{}subgroup}], [type](=direct, Cayley-operator, Cayley1-op, C-point, C1-p)
    {\it{}ddV}[3], {\it{}Curv}[3], // indexed by  [type](=direct, Cayley-operator, Cayley1-op)
    {\it{}focal}, {\it{}p},
    {\it{}focal\_l}, {\it{}focal\_u}, {\it{}focal\_v}; // parameters of parabola given by \(v = au\sp2 + bu + c\)

\nwused{\\{NW3OF8UE-1p0Y9w-1}}\nwidentuses{\\{{\nwixident{subgroup}}{subgroup}}\\{{\nwixident{v}}{v}}}\nwindexuse{\nwixident{subgroup}}{subgroup}{NW3OF8UE-2Cj0m9-4}\nwindexuse{\nwixident{v}}{v}{NW3OF8UE-2Cj0m9-4}\nwendcode{}\nwbegindocs{51}Here is the set of constants which allows to fine tune \MetaPost\ output
depending from the type of {\Tt{}\Rm{}{\it{}subgroup}\nwendquote} and {\Tt{}\Rm{}{\it{}metric}\nwendquote} used. The
quality of pictures will significantly depend from the number of
points chosen for iterations: too much lines will mess up the picture,
very few makes it incomplete or insensitive to the singular
regions. These numbers should be different for different combinations
of  {\Tt{}\Rm{}{\it{}subgroup}\nwendquote} and {\Tt{}\Rm{}{\it{}metric}\nwendquote}.
\nwenddocs{}\nwbegincode{52}\sublabel{NW3OF8UE-44zkoV-1}\nwmargintag{{\nwtagstyle{}\subpageref{NW3OF8UE-44zkoV-1}}}\moddef{Pictures tuning~{\nwtagstyle{}\subpageref{NW3OF8UE-44zkoV-1}}}\endmoddef\Rm{}\nwstartdeflinemarkup\nwusesondefline{\\{NW3OF8UE-1p0Y9w-1}}\nwenddeflinemarkup
{\bf{}const} {\bf{}int} {\it{}vilimits}[3][3] = {\nwlbrace}10, 20, 30,   // indexed by [{\it{}subgroup}], [{\it{}metric}]\nwindexdefn{\nwixident{vilimits}}{vilimits}{NW3OF8UE-44zkoV-1}
                    10, 10, 19,
                    10, 10, 10{\nwrbrace},
        {\it{}fsteps}[3][3] = {\nwlbrace}15, 15, 20,   // indexed by [{\it{}subgroup}], [{\it{}metric}]
                    15, 10, 20,
                    12, 15, 15{\nwrbrace};
{\bf{}float} {\it{}ulim} = 25, {\it{}vlim} = 25,\nwindexdefn{\nwixident{ulim}}{ulim}{NW3OF8UE-44zkoV-1}
        {\it{}flimits}[3][3] = {\nwlbrace}2.0, 2.0, 4.0,   // indexed by [{\it{}subgroup}], [{\it{}metric}]
                      10.0, 4.0, 4.0,
                      0.5, 0.5, 0.5{\nwrbrace},
        {\it{}vpoints}[3][10] = {\nwlbrace}0, 1.0\begin{math}\div\end{math}8, 1.0\begin{math}\div\end{math}4, 1.0\begin{math}\div\end{math}2, 1.0, 2.0, 3.0, 5.0, 8.0, 16.0,//[{\it{}metric}][{\it{}point}]
                       0, 1.0\begin{math}\div\end{math}8, 1.0\begin{math}\div\end{math}4, 1.0\begin{math}\div\end{math}2, 1, 2.0, 3.0, 6.0, 10.0, 20.0,
                       0,1.0\begin{math}\div\end{math}8, 1.0\begin{math}\div\end{math}4, 1.0\begin{math}\div\end{math}2, 1.0, 2.0, 3.0, 5.0, 10.0, 100{\nwrbrace};

\nwused{\\{NW3OF8UE-1p0Y9w-1}}\nwidentdefs{\\{{\nwixident{ulim}}{ulim}}\\{{\nwixident{vilimits}}{vilimits}}}\nwidentuses{\\{{\nwixident{metric}}{metric}}\\{{\nwixident{subgroup}}{subgroup}}}\nwindexuse{\nwixident{metric}}{metric}{NW3OF8UE-44zkoV-1}\nwindexuse{\nwixident{subgroup}}{subgroup}{NW3OF8UE-44zkoV-1}\nwendcode{}\nwbegindocs{53}Initialise the set of coordinates for a cycle
\nwenddocs{}\nwbegincode{54}\sublabel{NW3OF8UE-2mZmnQ-1}\nwmargintag{{\nwtagstyle{}\subpageref{NW3OF8UE-2mZmnQ-1}}}\moddef{Initialisation of coordinates~{\nwtagstyle{}\subpageref{NW3OF8UE-2mZmnQ-1}}}\endmoddef\Rm{}\nwstartdeflinemarkup\nwusesondefline{\\{NW3OF8UE-3Fz9v4-2}\\{NW3OF8UE-38hDUf-2}}\nwenddeflinemarkup
{\it{}init\_coord}(0);
{\it{}init\_coord}(1);
{\it{}put\_draw}(2);

\nwused{\\{NW3OF8UE-3Fz9v4-2}\\{NW3OF8UE-38hDUf-2}}\nwidentuses{\\{{\nwixident{init{\_}coord}}{init:uncoord}}\\{{\nwixident{put{\_}draw}}{put:undraw}}}\nwindexuse{\nwixident{init{\_}coord}}{init:uncoord}{NW3OF8UE-2mZmnQ-1}\nwindexuse{\nwixident{put{\_}draw}}{put:undraw}{NW3OF8UE-2mZmnQ-1}\nwendcode{}\nwbegindocs{55}\nwdocspar
\section{Symbolic Clifford Algebra Calculations}
\label{sec:cliff-algebra-calc}
This section finally starts to deal with Clifford algebras. We try to
make all possible calculations symbolically delaying the numeric
substitution to the latest stage. This produces a faster code as well.

\nwenddocs{}\nwbegindocs{56}\nwdocspar
\subsection{Initialisation of Clifford Numbers}
\label{sec:init-cliff-numb}

We initialise Clifford numbers first. Alternative ways to define
{\Tt{}\Rm{}{\it{}e}\nwendquote}, {\Tt{}\Rm{}{\it{}e0}\nwendquote} and {\Tt{}\Rm{}{\it{}e1}\nwendquote} are indicated in comments.
\nwenddocs{}\nwbegincode{57}\sublabel{NW3OF8UE-2xd863-1}\nwmargintag{{\nwtagstyle{}\subpageref{NW3OF8UE-2xd863-1}}}\moddef{Initialise Clifford numbers~{\nwtagstyle{}\subpageref{NW3OF8UE-2xd863-1}}}\endmoddef\Rm{}\nwstartdeflinemarkup\nwusesondefline{\\{NW3OF8UE-1p0Y9w-2}\\{NW3OF8UE-2KWnYy-1}}\nwprevnextdefs{\relax}{NW3OF8UE-2xd863-2}\nwenddeflinemarkup
{\it{}signum}  = {\bf{}numeric}({\it{}metric}-1); // the value of \(e\sb2\sp2\)
{\it{}M} = -1, 0,
    0, {\it{}signum};
//e = clifford\_unit(mu, indexed(M, symmetric2(),xi, psi));
{\it{}e} = {\it{}clifford\_unit}({\it{}mu}, {\it{}M});
{\it{}e0} = {\it{}e}.{\it{}subs}({\it{}mu} \begin{math}\equiv\end{math} 0);
{\it{}e1} = {\it{}e}.{\it{}subs}({\it{}mu} \begin{math}\equiv\end{math} 1);
// {\it{}e0} = {\it{}lst\_to\_clifford}({\bf{}lst}(1.0, 0), {\it{}mu}, {\it{}M});
// {\it{}e1} = {\it{}lst\_to\_clifford}({\bf{}lst}(0, 1.0), {\it{}mu}, {\it{}M});

\nwalsodefined{\\{NW3OF8UE-2xd863-2}\\{NW3OF8UE-2xd863-3}\\{NW3OF8UE-2xd863-4}}\nwused{\\{NW3OF8UE-1p0Y9w-2}\\{NW3OF8UE-2KWnYy-1}}\nwidentuses{\\{{\nwixident{metric}}{metric}}\\{{\nwixident{numeric}}{numeric}}}\nwindexuse{\nwixident{metric}}{metric}{NW3OF8UE-2xd863-1}\nwindexuse{\nwixident{numeric}}{numeric}{NW3OF8UE-2xd863-1}\nwendcode{}\nwbegindocs{58}Now we define matrices used for definition of the alternative Cayley
transforms~\cite{Kisil04b,Kisil05a}.
\nwenddocs{}\nwbegincode{59}\sublabel{NW3OF8UE-2xd863-2}\nwmargintag{{\nwtagstyle{}\subpageref{NW3OF8UE-2xd863-2}}}\moddef{Initialise Clifford numbers~{\nwtagstyle{}\subpageref{NW3OF8UE-2xd863-1}}}\plusendmoddef\Rm{}\nwstartdeflinemarkup\nwusesondefline{\\{NW3OF8UE-1p0Y9w-2}\\{NW3OF8UE-2KWnYy-1}}\nwprevnextdefs{NW3OF8UE-2xd863-1}{NW3OF8UE-2xd863-3}\nwenddeflinemarkup
{\it{}T} = {\it{}dirac\_ONE}(), {\it{}e0},  // Transformation to the alternative Cayley map
    {\it{}e0}, {\it{}dirac\_ONE}();   // is given by \(\begin{pmatrix} 1 \tab e\sb1 \cr e\sb1 \tab 1\end{pmatrix}\)
{\it{}TI} = {\it{}dirac\_ONE}(), -{\it{}e0},  // The inverse of {\it{}T}
    -{\it{}e0}, {\it{}dirac\_ONE}();   // is given by \(\begin{pmatrix} 1 \tab -e\sb1 \cr -e\sb1 \tab 1\end{pmatrix}\)

\nwused{\\{NW3OF8UE-1p0Y9w-2}\\{NW3OF8UE-2KWnYy-1}}\nwendcode{}\nwbegindocs{60}A form of matrix for the Cayley transform depends from the type of
metric.
\nwenddocs{}\nwbegincode{61}\sublabel{NW3OF8UE-2xd863-3}\nwmargintag{{\nwtagstyle{}\subpageref{NW3OF8UE-2xd863-3}}}\moddef{Initialise Clifford numbers~{\nwtagstyle{}\subpageref{NW3OF8UE-2xd863-1}}}\plusendmoddef\Rm{}\nwstartdeflinemarkup\nwusesondefline{\\{NW3OF8UE-1p0Y9w-2}\\{NW3OF8UE-2KWnYy-1}}\nwprevnextdefs{NW3OF8UE-2xd863-2}{NW3OF8UE-2xd863-4}\nwenddeflinemarkup
{\bf{}switch}  ({\it{}metric}) {\nwlbrace}
{\bf{}case} {\it{}elliptic}:
{\bf{}case} {\it{}hyperbolic}:
        {\it{}C} = {\it{}dirac\_ONE}(), -{\it{}e1},   // First Cayley transform \(\begin{pmatrix} 1 \tab -e\sb2 \cr \sigma e\sb2 \tab 1\end{pmatrix}\)
                {\it{}signum}\begin{math}\ast\end{math}{\it{}e1}, {\it{}dirac\_ONE}();
        {\it{}CI} = {\it{}dirac\_ONE}(), {\it{}e1},   // The inverse of {\it{}C}  \(\begin{pmatrix} 1 \tab e\sb2 \cr -\sigma e\sb2 \tab 1\end{pmatrix}\)
                -{\it{}signum}\begin{math}\ast\end{math}{\it{}e1}, {\it{}dirac\_ONE}();
        {\it{}C1} = {\it{}C}.{\it{}mul}({\it{}T}); // Second Cayley transform
        {\it{}C1I} = {\it{}TI}.{\it{}mul}({\it{}CI}); // The inverse of {\it{}C1}
        {\bf{}break};

\nwused{\\{NW3OF8UE-1p0Y9w-2}\\{NW3OF8UE-2KWnYy-1}}\nwidentuses{\\{{\nwixident{elliptic}}{elliptic}}\\{{\nwixident{hyperbolic}}{hyperbolic}}\\{{\nwixident{metric}}{metric}}}\nwindexuse{\nwixident{elliptic}}{elliptic}{NW3OF8UE-2xd863-3}\nwindexuse{\nwixident{hyperbolic}}{hyperbolic}{NW3OF8UE-2xd863-3}\nwindexuse{\nwixident{metric}}{metric}{NW3OF8UE-2xd863-3}\nwendcode{}\nwbegindocs{62}In the parabolic case there are two different (elliptic and
hyperbolic) types of the Cayley transform~\cite{Kisil04b,Kisil05a}.
\nwenddocs{}\nwbegincode{63}\sublabel{NW3OF8UE-2xd863-4}\nwmargintag{{\nwtagstyle{}\subpageref{NW3OF8UE-2xd863-4}}}\moddef{Initialise Clifford numbers~{\nwtagstyle{}\subpageref{NW3OF8UE-2xd863-1}}}\plusendmoddef\Rm{}\nwstartdeflinemarkup\nwusesondefline{\\{NW3OF8UE-1p0Y9w-2}\\{NW3OF8UE-2KWnYy-1}}\nwprevnextdefs{NW3OF8UE-2xd863-3}{\relax}\nwenddeflinemarkup
{\bf{}case} {\it{}parabolic}:
        {\it{}C} = {\it{}dirac\_ONE}(), -{\it{}e1}\begin{math}\ast\end{math}{\it{}half},  // First (elliptic) Cayley transform for the {\it{}parabolic} case
                -{\it{}e1}\begin{math}\ast\end{math}{\it{}half}, {\it{}dirac\_ONE}();
        {\it{}CI} = {\it{}dirac\_ONE}(), {\it{}e1}\begin{math}\ast\end{math}{\it{}half},  // The inverse of {\it{}C}
                {\it{}e1}\begin{math}\ast\end{math}{\it{}half}, {\it{}dirac\_ONE}();
        {\it{}C1} = {\it{}dirac\_ONE}(), -{\it{}e1}\begin{math}\ast\end{math}{\it{}half}, // Second (hyperbolic) Cayley transform for the {\it{}parabolic} case
                {\it{}e1}\begin{math}\ast\end{math}{\it{}half}, {\it{}dirac\_ONE}();
        {\it{}C1I} = {\it{}dirac\_ONE}(), {\it{}e1}\begin{math}\ast\end{math}{\it{}half}, // The inverse of {\it{}C1}
                -{\it{}e1}\begin{math}\ast\end{math}{\it{}half}, {\it{}dirac\_ONE}();
        {\bf{}break};
{\nwrbrace}

\nwused{\\{NW3OF8UE-1p0Y9w-2}\\{NW3OF8UE-2KWnYy-1}}\nwidentuses{\\{{\nwixident{elliptic}}{elliptic}}\\{{\nwixident{hyperbolic}}{hyperbolic}}\\{{\nwixident{parabolic}}{parabolic}}}\nwindexuse{\nwixident{elliptic}}{elliptic}{NW3OF8UE-2xd863-4}\nwindexuse{\nwixident{hyperbolic}}{hyperbolic}{NW3OF8UE-2xd863-4}\nwindexuse{\nwixident{parabolic}}{parabolic}{NW3OF8UE-2xd863-4}\nwendcode{}\nwbegindocs{64}\nwdocspar
\subsection{M\"obius Transformations}
\label{sec:mobi-transf-1}
We calculate all Moebius transformations along the orbits as well as
two their Cayley transforms only once in a symbolic way. Their usage
will be made through the \GiNaC\ substitution mechanism.

\nwenddocs{}\nwbegindocs{65} First, we define matrices {\Tt{}\Rm{}{\it{}Exp\_A}\nwendquote}, {\Tt{}\Rm{}{\it{}Exp\_N}\nwendquote},
{\Tt{}\Rm{}{\it{}Exp\_K}\nwendquote}~\cite[VI.1]{Lang85} related to the Iwasawa decomposition of
\(\SL\)~\cite[\S~III.1]{Lang85}.
\nwenddocs{}\nwbegincode{66}\sublabel{NW3OF8UE-1OFQOX-1}\nwmargintag{{\nwtagstyle{}\subpageref{NW3OF8UE-1OFQOX-1}}}\moddef{Calculation of Moebius transformations~{\nwtagstyle{}\subpageref{NW3OF8UE-1OFQOX-1}}}\endmoddef\Rm{}\nwstartdeflinemarkup\nwusesondefline{\\{NW3OF8UE-1p0Y9w-2}}\nwprevnextdefs{\relax}{NW3OF8UE-1OFQOX-2}\nwenddeflinemarkup
{\bf{}matrix} {\it{}Exp\_A}(2, 2), {\it{}Exp\_N}(2, 2), {\it{}Exp\_K}(2, 2);
{\it{}Exp\_A} = {\it{}exp}({\it{}t}) \begin{math}\ast\end{math} {\it{}dirac\_ONE}(), 0, // Matrix \(\begin{pmatrix} e\sp{t} \tab0\cr 0\tab e\sp{-t}\end{pmatrix} = \exp \begin{pmatrix}1\tab0\cr 0\tab -1\end{pmatrix} \)
                    0, {\it{}exp}(-{\it{}t}) \begin{math}\ast\end{math} {\it{}dirac\_ONE}();
{\it{}Exp\_N} = {\it{}dirac\_ONE}(), {\it{}t} \begin{math}\ast\end{math} {\it{}e0}, // Matrix \(  \begin{pmatrix}0\tab t\cr 0\tab0\end{pmatrix}  =\exp\begin{pmatrix}0\tab1\cr 0\tab0\end{pmatrix} \)
                    0, {\it{}dirac\_ONE}();
{\it{}Exp\_K} = {\it{}cos}({\it{}t}) \begin{math}\ast\end{math} {\it{}dirac\_ONE}(), {\it{}sin}({\it{}t}) \begin{math}\ast\end{math} {\it{}e0}, // Matrix \( \begin{pmatrix}\cos t \tab -\sin t\cr \sin t\tab \cos t \end{pmatrix} = \exp\begin{pmatrix}0\tab-1\cr 1\tab0 \end{pmatrix} \)
                   {\it{}sin}({\it{}t}) \begin{math}\ast\end{math} {\it{}e0}, {\it{}cos}({\it{}t}) \begin{math}\ast\end{math} {\it{}dirac\_ONE}();
{\bf{}ex} {\it{}Exp}[3] = {\nwlbrace}{\it{}Exp\_A}, {\it{}Exp\_N}, {\it{}Exp\_K}{\nwrbrace};
{\bf{}matrix} {\it{}E}(2, 2);

\nwalsodefined{\\{NW3OF8UE-1OFQOX-2}}\nwused{\\{NW3OF8UE-1p0Y9w-2}}\nwendcode{}\nwbegindocs{67}Here we symbolically calculate the related
M\"obius transformations, as well as its two images under Cayley transform
{\Tt{}\Rm{}{\it{}C}\nwendquote} and {\Tt{}\Rm{}{\it{}C1}\nwendquote} for operators and finally two Cayley images for
points.
\nwenddocs{}\nwbegincode{68}\sublabel{NW3OF8UE-1OFQOX-2}\nwmargintag{{\nwtagstyle{}\subpageref{NW3OF8UE-1OFQOX-2}}}\moddef{Calculation of Moebius transformations~{\nwtagstyle{}\subpageref{NW3OF8UE-1OFQOX-1}}}\plusendmoddef\Rm{}\nwstartdeflinemarkup\nwusesondefline{\\{NW3OF8UE-1p0Y9w-2}}\nwprevnextdefs{NW3OF8UE-1OFQOX-1}{\relax}\nwenddeflinemarkup
{\bf{}for} ({\it{}subgroup} = {\it{}subgroup\_A}; {\it{}subgroup} \begin{math}\leq\end{math} {\it{}subgroup\_K}; {\it{}subgroup}\protect\PP) {\nwlbrace}
        {\bf{}try} {\nwlbrace}
                {\it{}Moebius}[{\it{}subgroup}][0] = {\it{}clifford\_moebius\_map}({\it{}Exp}[{\it{}subgroup}], {\bf{}lst}({\it{}x}, {\it{}y}), {\it{}M});
                {\commopen} Cayley transforms of operators {\commclose}
                {\it{}Moebius}[{\it{}subgroup}][1] = {\it{}clifford\_moebius\_map}({\it{}canonicalize\_clifford}(
                            {\it{}C}.{\it{}mul}({\it{}ex\_to}\begin{math}<\end{math}{\bf{}matrix}\begin{math}>\end{math}({\it{}Exp}[{\it{}subgroup}]).{\it{}mul}({\it{}CI}))), {\bf{}lst}({\it{}x}, {\it{}y}), {\it{}M});
                {\it{}Moebius}[{\it{}subgroup}][2] = {\it{}clifford\_moebius\_map}({\it{}canonicalize\_clifford}(
                            {\it{}C1}.{\it{}mul}({\it{}ex\_to}\begin{math}<\end{math}{\bf{}matrix}\begin{math}>\end{math}({\it{}Exp}[{\it{}subgroup}]).{\it{}mul}({\it{}C1I}))), {\bf{}lst}({\it{}x}, {\it{}y}), {\it{}M});
                {\commopen} Cayley transforms of points {\commclose}
                {\it{}Moebius}[{\it{}subgroup}][3] = {\it{}clifford\_moebius\_map}({\it{}C}.{\it{}mul}({\it{}ex\_to}\begin{math}<\end{math}{\bf{}matrix}\begin{math}>\end{math}({\it{}Exp}[{\it{}subgroup}])), {\bf{}lst}({\it{}x}, {\it{}y}), {\it{}M});
                {\it{}Moebius}[{\it{}subgroup}][4] = {\it{}clifford\_moebius\_map}({\it{}canonicalize\_clifford}(
                            {\it{}C1}.{\it{}mul}({\it{}ex\_to}\begin{math}<\end{math}{\bf{}matrix}\begin{math}>\end{math}({\it{}Exp}[{\it{}subgroup}]))), {\bf{}lst}({\it{}x}, {\it{}y}), {\it{}M});
        {\nwrbrace} {\bf{}catch}  ({\it{}exception} &{\it{}p}) {\nwlbrace}
                {\it{}cerr} \begin{math}\ll\end{math} {\tt{}"*** Got problem in vector fields: "} \begin{math}\ll\end{math} {\it{}p}.{\it{}what}() \begin{math}\ll\end{math} {\it{}endl};
        {\nwrbrace}
{\nwrbrace}

\nwused{\\{NW3OF8UE-1p0Y9w-2}}\nwidentuses{\\{{\nwixident{catch}}{catch}}\\{{\nwixident{subgroup}}{subgroup}}\\{{\nwixident{subgroup{\_}A}}{subgroup:unA}}\\{{\nwixident{subgroup{\_}K}}{subgroup:unK}}}\nwindexuse{\nwixident{catch}}{catch}{NW3OF8UE-1OFQOX-2}\nwindexuse{\nwixident{subgroup}}{subgroup}{NW3OF8UE-1OFQOX-2}\nwindexuse{\nwixident{subgroup{\_}A}}{subgroup:unA}{NW3OF8UE-1OFQOX-2}\nwindexuse{\nwixident{subgroup{\_}K}}{subgroup:unK}{NW3OF8UE-1OFQOX-2}\nwendcode{}\nwbegindocs{69}\nwdocspar
\subsection{Symbolic Calculations of the Vector Fields}
\label{sec:symb-calc-vect}

We calculate symbolic expressions for the vector fields of three
subgroups \(A\), \(N\) and \(K\),  which are stated in
\cite[Lemmas~\ref{E-le:an-action} and~\ref{E-le:k-action}]{Kisil05a} and
shown on \cite[Figures~\ref{E-fig:a-n-action}
and~\ref{E-fig:k-sungroup}]{Kisil05a}. The formula for a derived
representation \(\rho(X)\) of a vector field \(X\) used here
is~\cite[\S~VI.1]{Lang85}:
\begin{displaymath}
  \rho(X) = \frac{d}{dt}\rho(e^{tX})\such_{t=0}.
\end{displaymath}
Results of calculations are directed to {\Tt{}\Rm{}{\it{}stdout}\nwendquote} and will be used
for a better drawing of orbits, see~\ref{sec:defines}.

We calculate {\Tt{}\Rm{}{\it{}Jacob}\nwendquote}ian of the M\"obius transformations in order
to suply to \MetaPost\ the tangents of the transverse lines.

\nwenddocs{}\nwbegincode{70}\sublabel{NW3OF8UE-4BpIv1-1}\nwmargintag{{\nwtagstyle{}\subpageref{NW3OF8UE-4BpIv1-1}}}\moddef{Calculation of vector fields~{\nwtagstyle{}\subpageref{NW3OF8UE-4BpIv1-1}}}\endmoddef\Rm{}\nwstartdeflinemarkup\nwusesondefline{\\{NW3OF8UE-1p0Y9w-2}}\nwprevnextdefs{\relax}{NW3OF8UE-4BpIv1-2}\nwenddeflinemarkup
{\bf{}try} {\nwlbrace}
    {\it{}cout} \begin{math}\ll\end{math} {\tt{}"Vect field {\char92}t Direct {\char92}t{\char92}t In Cayley {\char92}t{\char92}t In Cayley1"} \begin{math}\ll\end{math} {\it{}endl};
    {\bf{}for} ({\it{}subgroup} = {\it{}subgroup\_A}; {\it{}subgroup} \begin{math}\leq\end{math} {\it{}subgroup\_K}; {\it{}subgroup}\protect\PP) {\nwlbrace}
        {\bf{}for} ({\bf{}int} {\it{}cayley} = 0; {\it{}cayley} \begin{math}<\end{math} 3 ; {\it{}cayley}\protect\PP) {\nwlbrace}
            {\bf{}lst} {\it{}Moeb} = {\it{}ex\_to}\begin{math}<\end{math}{\bf{}lst}\begin{math}>\end{math}({\it{}Moebius}[{\it{}subgroup}][{\it{}cayley}?{\it{}cayley}+2:0]);
            {\it{}dV}[{\it{}subgroup}][{\it{}cayley}] = {\it{}Moebius}[{\it{}subgroup}][{\it{}cayley}].{\it{}diff}({\it{}t}).{\it{}subs}({\it{}t} \begin{math}\equiv\end{math} 0);
            {\it{}Jacob}[{\it{}subgroup}][{\it{}cayley}] = {\bf{}matrix}(2,2, {\bf{}lst}({\it{}Moeb}.{\it{}op}(0).{\it{}diff}({\it{}x}), {\it{}Moeb}.{\it{}op}(0).{\it{}diff}({\it{}y}),
               {\it{}Moeb}.{\it{}op}(1).{\it{}diff}({\it{}x}), {\it{}Moeb}.{\it{}op}(1).{\it{}diff}({\it{}y})));
            // Transformation of a direction by a Jacobian
            {\bf{}for} ({\bf{}int} {\it{}i}=0; {\it{}i} \begin{math}<\end{math} 2; {\it{}i}\protect\PP)
                {\it{}trans\_dir}[{\it{}subgroup}][{\it{}cayley}] = {\it{}Jacob}[{\it{}subgroup}][{\it{}cayley}].{\it{}mul}({\bf{}matrix}(2, 1, {\bf{}lst}({\it{}tr\_u}, {\it{}tr\_v})));
        {\nwrbrace}
        {\it{}cout} \begin{math}\ll\end{math} {\tt{}"  d"} \begin{math}\ll\end{math} {\it{}sgroup}[{\it{}subgroup}] \begin{math}\ll\end{math} {\tt{}" is:{\char92}t"} \begin{math}\ll\end{math} {\it{}dV}[{\it{}subgroup}][0]
             \begin{math}\ll\end{math} {\tt{}";{\char92}t"} \begin{math}\ll\end{math} {\it{}dV}[{\it{}subgroup}][1] \begin{math}\ll\end{math} {\tt{}";{\char92}t "} \begin{math}\ll\end{math} {\it{}dV}[{\it{}subgroup}][2] \begin{math}\ll\end{math} {\it{}endl};
{\nwrbrace}

\nwalsodefined{\\{NW3OF8UE-4BpIv1-2}}\nwused{\\{NW3OF8UE-1p0Y9w-2}}\nwidentuses{\\{{\nwixident{sgroup}}{sgroup}}\\{{\nwixident{subgroup}}{subgroup}}\\{{\nwixident{subgroup{\_}A}}{subgroup:unA}}\\{{\nwixident{subgroup{\_}K}}{subgroup:unK}}}\nwindexuse{\nwixident{sgroup}}{sgroup}{NW3OF8UE-4BpIv1-1}\nwindexuse{\nwixident{subgroup}}{subgroup}{NW3OF8UE-4BpIv1-1}\nwindexuse{\nwixident{subgroup{\_}A}}{subgroup:unA}{NW3OF8UE-4BpIv1-1}\nwindexuse{\nwixident{subgroup{\_}K}}{subgroup:unK}{NW3OF8UE-4BpIv1-1}\nwendcode{}\nwbegindocs{71}We also calculate curvature of the orbits using the formula~\cite[\S~5.1(20)]{DubrovinNovikovFomenkoI}
\begin{displaymath}
  k = \frac{\modulus{\ddot{x}\dot{y}-\ddot{y}\dot{x}}}{(\dot{x}^2+\dot{y}^2)^{3/2}}
\end{displaymath}

\nwenddocs{}\nwbegincode{72}\sublabel{NW3OF8UE-4BpIv1-2}\nwmargintag{{\nwtagstyle{}\subpageref{NW3OF8UE-4BpIv1-2}}}\moddef{Calculation of vector fields~{\nwtagstyle{}\subpageref{NW3OF8UE-4BpIv1-1}}}\plusendmoddef\Rm{}\nwstartdeflinemarkup\nwusesondefline{\\{NW3OF8UE-1p0Y9w-2}}\nwprevnextdefs{NW3OF8UE-4BpIv1-1}{\relax}\nwenddeflinemarkup
    {\bf{}for} ({\bf{}int} {\it{}cayley} = 0; {\it{}cayley} \begin{math}<\end{math} 3 ; {\it{}cayley}\protect\PP) {\nwlbrace}
        {\it{}ddV}[{\it{}cayley}] = {\it{}Moebius}[{\it{}subgroup\_K}][{\it{}cayley}].{\it{}diff}({\it{}t}, 2).{\it{}subs}({\it{}t} \begin{math}\equiv\end{math} 0);
        {\bf{}ex} {\it{}du} = {\it{}ex\_to}\begin{math}<\end{math}{\bf{}lst}\begin{math}>\end{math}({\it{}dV}[{\it{}subgroup\_K}][{\it{}cayley}]).{\it{}op}(0);
        {\bf{}ex} {\it{}dv} = {\it{}ex\_to}\begin{math}<\end{math}{\bf{}lst}\begin{math}>\end{math}({\it{}dV}[{\it{}subgroup\_K}][{\it{}cayley}]).{\it{}op}(1);
        {\bf{}ex} {\it{}ddu} = {\it{}ex\_to}\begin{math}<\end{math}{\bf{}lst}\begin{math}>\end{math}({\it{}ddV}[{\it{}cayley}]).{\it{}op}(0);
        {\bf{}ex} {\it{}ddv} = {\it{}ex\_to}\begin{math}<\end{math}{\bf{}lst}\begin{math}>\end{math}({\it{}ddV}[{\it{}cayley}]).{\it{}op}(1);
        {\it{}Curv}[{\it{}cayley}] = {\it{}normal}(({\it{}ddu} \begin{math}\ast\end{math} {\it{}dv} - {\it{}du} \begin{math}\ast\end{math} {\it{}ddv})\begin{math}\div\end{math}{\it{}pow}({\it{}du}\begin{math}\ast\end{math}{\it{}du}+{\it{}dv}\begin{math}\ast\end{math}{\it{}dv}, 1.5));
    {\nwrbrace}
    {\it{}cout} \begin{math}\ll\end{math} {\tt{}"Curvature of K-orbits on the v-axis: "}
         \begin{math}\ll\end{math} {\it{}normal}({\it{}Curv}[0].{\it{}subs}({\it{}x} \begin{math}\equiv\end{math} 0)) \begin{math}\ll\end{math} {\it{}endl};

{\nwrbrace} {\bf{}catch}  ({\it{}exception} &{\it{}p}) {\nwlbrace}\nwindexdefn{\nwixident{catch}}{catch}{NW3OF8UE-4BpIv1-2}
    {\it{}cerr} \begin{math}\ll\end{math} {\tt{}"*** Got problem in vector fields: "} \begin{math}\ll\end{math} {\it{}p}.{\it{}what}() \begin{math}\ll\end{math} {\it{}endl};
{\nwrbrace}

\nwused{\\{NW3OF8UE-1p0Y9w-2}}\nwidentdefs{\\{{\nwixident{catch}}{catch}}}\nwidentuses{\\{{\nwixident{subgroup{\_}K}}{subgroup:unK}}\\{{\nwixident{v}}{v}}}\nwindexuse{\nwixident{subgroup{\_}K}}{subgroup:unK}{NW3OF8UE-4BpIv1-2}\nwindexuse{\nwixident{v}}{v}{NW3OF8UE-4BpIv1-2}\nwendcode{}\nwbegindocs{73}\nwdocspar
\section{Numeric Calculations with Clifford Algebras}
\label{sec:numer-calc-with}
Numeric calculations are done in to fashions:
\begin{enumerate}
\item Through a substitution of  numeric values to symbols in some previous
  symbolic results, subsection~\ref{sec:numer-calc-orbits}
  and~\ref{sec:cayl-transf-imag};
\item Direct calculations with numeric \GiNaC\ classes,
  subsection~\ref{sec:future-past-transf}.
\end{enumerate}

\nwenddocs{}\nwbegindocs{74}\nwdocspar
The first example of substituion approach is the drawing of vector fields.
Three vector fields are drawn by arrows into a \MetaPost\ file.
\nwenddocs{}\nwbegincode{75}\sublabel{NW3OF8UE-3M9GDT-1}\nwmargintag{{\nwtagstyle{}\subpageref{NW3OF8UE-3M9GDT-1}}}\moddef{Drawing arrows~{\nwtagstyle{}\subpageref{NW3OF8UE-3M9GDT-1}}}\endmoddef\Rm{}\nwstartdeflinemarkup\nwusesondefline{\\{NW3OF8UE-1p0Y9w-3}}\nwenddeflinemarkup
{\it{}fileout}[0] = {\it{}openfile}({\tt{}"arrows"}); // open \MetaPost\  file
{\it{}color\_grade} = {\it{}grey};
{\bf{}for} ({\bf{}int} {\it{}k} = -10; {\it{}k} \begin{math}<\end{math} 10; {\it{}k}\protect\PP) // cycle over horizontal dir
    {\bf{}for} ({\bf{}int} {\it{}j} = 0; {\it{}j} \begin{math}<\end{math} 11; {\it{}j}\protect\PP) {\nwlbrace} // cycle over vertical dir
        {\it{}u} = {\it{}k}\begin{math}\div\end{math}3.0;   // Calculate coordinates of the point
        {\it{}v} = {\it{}j}\begin{math}\div\end{math}3.0;
        {\it{}u\_res} = {\it{}ex\_to}\begin{math}<\end{math}{\bf{}numeric}\begin{math}>\end{math}({\it{}dV}[{\it{}subgroup}][0].{\it{}subs}({\bf{}lst}({\it{}x} \begin{math}\equiv\end{math}  {\it{}u}, {\it{}y} \begin{math}\equiv\end{math} {\it{}v})).{\it{}op}(0)); // Get numeric
        {\it{}v\_res} = {\it{}ex\_to}\begin{math}<\end{math}{\bf{}numeric}\begin{math}>\end{math}({\it{}dV}[{\it{}subgroup}][0].{\it{}subs}({\bf{}lst}({\it{}x} \begin{math}\equiv\end{math}  {\it{}u}, {\it{}y} \begin{math}\equiv\end{math} {\it{}v})).{\it{}op}(1));
        {\it{}fprintf}({\it{}fileout}[0], {\tt{}"myarrow ((a
                {\it{}u}, {\it{}v}, {\it{}u}, {\it{}u\_res}.{\it{}to\_double}(), {\it{}v}, {\it{}v\_res}.{\it{}to\_double}());
        {\it{}fprintf}({\it{}fileout}[0], {\tt{}" withcolor  
    {\nwrbrace}
{\it{}fclose}({\it{}fileout}[0]);

\nwused{\\{NW3OF8UE-1p0Y9w-3}}\nwidentuses{\\{{\nwixident{color{\_}name}}{color:unname}}\\{{\nwixident{fileout}}{fileout}}\\{{\nwixident{grey}}{grey}}\\{{\nwixident{numeric}}{numeric}}\\{{\nwixident{openfile}}{openfile}}\\{{\nwixident{subgroup}}{subgroup}}\\{{\nwixident{u}}{u}}\\{{\nwixident{v}}{v}}}\nwindexuse{\nwixident{color{\_}name}}{color:unname}{NW3OF8UE-3M9GDT-1}\nwindexuse{\nwixident{fileout}}{fileout}{NW3OF8UE-3M9GDT-1}\nwindexuse{\nwixident{grey}}{grey}{NW3OF8UE-3M9GDT-1}\nwindexuse{\nwixident{numeric}}{numeric}{NW3OF8UE-3M9GDT-1}\nwindexuse{\nwixident{openfile}}{openfile}{NW3OF8UE-3M9GDT-1}\nwindexuse{\nwixident{subgroup}}{subgroup}{NW3OF8UE-3M9GDT-1}\nwindexuse{\nwixident{u}}{u}{NW3OF8UE-3M9GDT-1}\nwindexuse{\nwixident{v}}{v}{NW3OF8UE-3M9GDT-1}\nwendcode{}\nwbegindocs{76}\nwdocspar
\subsection{Numeric Calculations of Orbits and Transverses}
\label{sec:numer-calc-orbits}
For any of three possibility $e_2^2=-1$, $0$, $1$ and three
possible subgroups (\(A\), \(N\), \(K\)) orbits are constructed. First we open output
\MetaPost\ files.
\nwenddocs{}\nwbegincode{77}\sublabel{NW3OF8UE-3Fz9v4-1}\nwmargintag{{\nwtagstyle{}\subpageref{NW3OF8UE-3Fz9v4-1}}}\moddef{Building orbits~{\nwtagstyle{}\subpageref{NW3OF8UE-3Fz9v4-1}}}\endmoddef\Rm{}\nwstartdeflinemarkup\nwusesondefline{\\{NW3OF8UE-1p0Y9w-3}}\nwprevnextdefs{\relax}{NW3OF8UE-3Fz9v4-2}\nwenddeflinemarkup
{\it{}direct} = {\bf{}true};
{\it{}fileout}[0] = {\it{}openfile}({\tt{}"orbit"});
{\it{}fileout}[1] = {\it{}openfile}({\tt{}"cayley"}); // Cayley transform of the orbits
{\it{}fileout}[2] = {\it{}openfile}({\tt{}"cayl-a"}); // Alternative Cayley transform of the orbits

\nwalsodefined{\\{NW3OF8UE-3Fz9v4-2}}\nwused{\\{NW3OF8UE-1p0Y9w-3}}\nwidentuses{\\{{\nwixident{fileout}}{fileout}}\\{{\nwixident{openfile}}{openfile}}}\nwindexuse{\nwixident{fileout}}{fileout}{NW3OF8UE-3Fz9v4-1}\nwindexuse{\nwixident{openfile}}{openfile}{NW3OF8UE-3Fz9v4-1}\nwendcode{}\nwbegindocs{78} This chunk runs iterations over the different orbits, which are
initiated by the point {\Tt{}\Rm{}{\it{}vi}\nwendquote}.
\nwenddocs{}\nwbegincode{79}\sublabel{NW3OF8UE-3Fz9v4-2}\nwmargintag{{\nwtagstyle{}\subpageref{NW3OF8UE-3Fz9v4-2}}}\moddef{Building orbits~{\nwtagstyle{}\subpageref{NW3OF8UE-3Fz9v4-1}}}\plusendmoddef\Rm{}\nwstartdeflinemarkup\nwusesondefline{\\{NW3OF8UE-1p0Y9w-3}}\nwprevnextdefs{NW3OF8UE-3Fz9v4-1}{\relax}\nwenddeflinemarkup
{\bf{}for} ({\bf{}int} {\it{}vi} = 0; {\it{}vi} \begin{math}<\end{math} {\it{}vilimits}[{\it{}subgroup}][{\it{}metric}]; {\it{}vi}\protect\PP) {\nwlbrace} // iterator over orbits
        {\it{}color\_grade} = 1.2\begin{math}\ast\end{math}{\it{}vi}\begin{math}\div\end{math}{\it{}vilimits}[{\it{}subgroup}][{\it{}metric}];
    {\bf{}if} ({\it{}subgroup} \begin{math}\equiv\end{math} {\it{}subgroup\_K})
            {\it{}cout} \begin{math}\ll\end{math} {\it{}formula}[{\it{}metric}] ;
    \LA{}Initialisation of coordinates~{\nwtagstyle{}\subpageref{NW3OF8UE-2mZmnQ-1}}\RA{}
    \LA{}Nodes iterations~{\nwtagstyle{}\subpageref{NW3OF8UE-bYD0P-1}}\RA{}
    \LA{}Close all curves~{\nwtagstyle{}\subpageref{NW3OF8UE-Ix3jQ-1}}\RA{}
    \LA{}Check parabolas~{\nwtagstyle{}\subpageref{NW3OF8UE-2qDOKC-1}}\RA{}
{\nwrbrace}
\LA{}Closing all files~{\nwtagstyle{}\subpageref{NW3OF8UE-nrvMS-1}}\RA{}

\nwused{\\{NW3OF8UE-1p0Y9w-3}}\nwidentuses{\\{{\nwixident{metric}}{metric}}\\{{\nwixident{subgroup}}{subgroup}}\\{{\nwixident{subgroup{\_}K}}{subgroup:unK}}\\{{\nwixident{vilimits}}{vilimits}}}\nwindexuse{\nwixident{metric}}{metric}{NW3OF8UE-3Fz9v4-2}\nwindexuse{\nwixident{subgroup}}{subgroup}{NW3OF8UE-3Fz9v4-2}\nwindexuse{\nwixident{subgroup{\_}K}}{subgroup:unK}{NW3OF8UE-3Fz9v4-2}\nwindexuse{\nwixident{vilimits}}{vilimits}{NW3OF8UE-3Fz9v4-2}\nwendcode{}\nwbegindocs{80}Each orbit is processed by iteration over the ``time'' parameter
{\Tt{}\Rm{}{\it{}j}\nwendquote} on the orbit. For each node on an orbit an entry is put into
appropriate \MetaPost\ file, formulae from papers are numerically
checked and all Cayley transforms are produced.

\nwenddocs{}\nwbegincode{81}\sublabel{NW3OF8UE-bYD0P-1}\nwmargintag{{\nwtagstyle{}\subpageref{NW3OF8UE-bYD0P-1}}}\moddef{Nodes iterations~{\nwtagstyle{}\subpageref{NW3OF8UE-bYD0P-1}}}\endmoddef\Rm{}\nwstartdeflinemarkup\nwusesondefline{\\{NW3OF8UE-3Fz9v4-2}}\nwenddeflinemarkup
{\bf{}for} ({\bf{}int} {\it{}j} = -{\it{}fsteps}[{\it{}subgroup}][{\it{}metric}]; {\it{}j} \begin{math}\leq\end{math} {\it{}fsteps}[{\it{}subgroup}][{\it{}metric}]; {\it{}j}\protect\PP ) {\nwlbrace}
        {\bf{}float} {\it{}f} = {\it{}flimits}[{\it{}subgroup}][{\it{}metric}]\begin{math}\ast\end{math}{\it{}j}\begin{math}\div\end{math}{\it{}fsteps}[{\it{}subgroup}][{\it{}metric}]; // the angle of rotation
        \LA{}Generating one entry~{\nwtagstyle{}\subpageref{NW3OF8UE-4P0NfT-1}}\RA{}
        \LA{}Check formulas in the paper~{\nwtagstyle{}\subpageref{NW3OF8UE-47Sd20-1}}\RA{}
        \LA{}Producing Cayley transform of the orbit~{\nwtagstyle{}\subpageref{NW3OF8UE-D1caK-1}}\RA{}
    {\nwrbrace}

\nwused{\\{NW3OF8UE-3Fz9v4-2}}\nwidentuses{\\{{\nwixident{metric}}{metric}}\\{{\nwixident{subgroup}}{subgroup}}}\nwindexuse{\nwixident{metric}}{metric}{NW3OF8UE-bYD0P-1}\nwindexuse{\nwixident{subgroup}}{subgroup}{NW3OF8UE-bYD0P-1}\nwendcode{}\nwbegindocs{82}Closing all files when finishing drawing orbits or transverses
\nwenddocs{}\nwbegincode{83}\sublabel{NW3OF8UE-nrvMS-1}\nwmargintag{{\nwtagstyle{}\subpageref{NW3OF8UE-nrvMS-1}}}\moddef{Closing all files~{\nwtagstyle{}\subpageref{NW3OF8UE-nrvMS-1}}}\endmoddef\Rm{}\nwstartdeflinemarkup\nwusesondefline{\\{NW3OF8UE-3Fz9v4-2}\\{NW3OF8UE-38hDUf-2}}\nwenddeflinemarkup
{\it{}fclose}({\it{}fileout}[0]);
{\it{}fclose}({\it{}fileout}[1]);
{\it{}fclose}({\it{}fileout}[2]);

\nwused{\\{NW3OF8UE-3Fz9v4-2}\\{NW3OF8UE-38hDUf-2}}\nwidentuses{\\{{\nwixident{fileout}}{fileout}}}\nwindexuse{\nwixident{fileout}}{fileout}{NW3OF8UE-nrvMS-1}\nwendcode{}\nwbegindocs{84}\nwdocspar
\subsection{Building of Transverses}
\label{sec:building-transverses}

Construction of transverses to the orbits follows the same structure
as for orbits themselves with the changed order of iterations over
time parameter and orbit origins. We again start from opening of the
corresponding files.
\nwenddocs{}\nwbegincode{85}\sublabel{NW3OF8UE-38hDUf-1}\nwmargintag{{\nwtagstyle{}\subpageref{NW3OF8UE-38hDUf-1}}}\moddef{Building transverses~{\nwtagstyle{}\subpageref{NW3OF8UE-38hDUf-1}}}\endmoddef\Rm{}\nwstartdeflinemarkup\nwusesondefline{\\{NW3OF8UE-1p0Y9w-3}}\nwprevnextdefs{\relax}{NW3OF8UE-38hDUf-2}\nwenddeflinemarkup
{\it{}direct} = {\bf{}false};
{\it{}fileout}[0] = {\it{}openfile}({\tt{}"orbit-t"});
{\it{}fileout}[1] = {\it{}openfile}({\tt{}"cayley-t"}); // Cayley transform of transverses
{\it{}fileout}[2] = {\it{}openfile}({\tt{}"cayl-a-t"}); // Alternative Cayley transform of transverses
{\it{}color\_grade} = 1.2;
\LA{}Define transverse directions~{\nwtagstyle{}\subpageref{NW3OF8UE-4LL0eE-1}}\RA{};

\nwalsodefined{\\{NW3OF8UE-38hDUf-2}}\nwused{\\{NW3OF8UE-1p0Y9w-3}}\nwidentuses{\\{{\nwixident{fileout}}{fileout}}\\{{\nwixident{openfile}}{openfile}}}\nwindexuse{\nwixident{fileout}}{fileout}{NW3OF8UE-38hDUf-1}\nwindexuse{\nwixident{openfile}}{openfile}{NW3OF8UE-38hDUf-1}\nwendcode{}\nwbegindocs{86}Thus chunk performs iterations over the transverse lines.
\nwenddocs{}\nwbegincode{87}\sublabel{NW3OF8UE-38hDUf-2}\nwmargintag{{\nwtagstyle{}\subpageref{NW3OF8UE-38hDUf-2}}}\moddef{Building transverses~{\nwtagstyle{}\subpageref{NW3OF8UE-38hDUf-1}}}\plusendmoddef\Rm{}\nwstartdeflinemarkup\nwusesondefline{\\{NW3OF8UE-1p0Y9w-3}}\nwprevnextdefs{NW3OF8UE-38hDUf-1}{\relax}\nwenddeflinemarkup
{\bf{}for} ({\bf{}int} {\it{}j} = -{\it{}fsteps}[{\it{}subgroup}][{\it{}metric}]; {\it{}j} \begin{math}\leq\end{math} {\it{}fsteps}[{\it{}subgroup}][{\it{}metric}]; {\it{}j}\protect\PP ) {\nwlbrace}
    {\bf{}float} {\it{}f} = {\it{}flimits}[{\it{}subgroup}][{\it{}metric}]\begin{math}\ast\end{math}{\it{}j}\begin{math}\div\end{math}{\it{}fsteps}[{\it{}subgroup}][{\it{}metric}];// the angle of rotation
    \LA{}Initialisation of coordinates~{\nwtagstyle{}\subpageref{NW3OF8UE-2mZmnQ-1}}\RA{}
    {\bf{}for} ({\bf{}int} {\it{}vi} = 0; {\it{}vi} \begin{math}<\end{math} {\it{}vilimits}[{\it{}subgroup}][{\it{}metric}]; {\it{}vi}\protect\PP) {\nwlbrace} // iterator over orbits
        {\it{}vval} = {\it{}vpoints}[{\it{}metric}][{\it{}vi}];
        \LA{}Generating one entry~{\nwtagstyle{}\subpageref{NW3OF8UE-4P0NfT-1}}\RA{}
        \LA{}Producing Cayley transform of the orbit~{\nwtagstyle{}\subpageref{NW3OF8UE-D1caK-1}}\RA{}
    {\nwrbrace}
    \LA{}Close all curves~{\nwtagstyle{}\subpageref{NW3OF8UE-Ix3jQ-1}}\RA{}
{\nwrbrace}
\LA{}Closing all files~{\nwtagstyle{}\subpageref{NW3OF8UE-nrvMS-1}}\RA{}

\nwused{\\{NW3OF8UE-1p0Y9w-3}}\nwidentuses{\\{{\nwixident{metric}}{metric}}\\{{\nwixident{subgroup}}{subgroup}}\\{{\nwixident{vilimits}}{vilimits}}}\nwindexuse{\nwixident{metric}}{metric}{NW3OF8UE-38hDUf-2}\nwindexuse{\nwixident{subgroup}}{subgroup}{NW3OF8UE-38hDUf-2}\nwindexuse{\nwixident{vilimits}}{vilimits}{NW3OF8UE-38hDUf-2}\nwendcode{}\nwbegindocs{88}All \MetaPost\  \texttt{draw} statements should be closed  at the
end of run.
\nwenddocs{}\nwbegincode{89}\sublabel{NW3OF8UE-Ix3jQ-1}\nwmargintag{{\nwtagstyle{}\subpageref{NW3OF8UE-Ix3jQ-1}}}\moddef{Close all curves~{\nwtagstyle{}\subpageref{NW3OF8UE-Ix3jQ-1}}}\endmoddef\Rm{}\nwstartdeflinemarkup\nwusesondefline{\\{NW3OF8UE-3Fz9v4-2}\\{NW3OF8UE-38hDUf-2}}\nwenddeflinemarkup
{\it{}close\_curve}(0);
{\it{}close\_curve}(1);
{\it{}close\_curve}(2);

\nwused{\\{NW3OF8UE-3Fz9v4-2}\\{NW3OF8UE-38hDUf-2}}\nwidentuses{\\{{\nwixident{close{\_}curve}}{close:uncurve}}}\nwindexuse{\nwixident{close{\_}curve}}{close:uncurve}{NW3OF8UE-Ix3jQ-1}\nwendcode{}\nwbegindocs{90}\nwdocspar
\subsection{Future-to-Past Transformations}
\label{sec:future-past-transf}
We finish our code by producing a set of frames of transformations
from future to past of light cone. First we create a necessary
{\Tt{}\Rm{}{\it{}hyperbolic}\nwendquote} setup and use {\Tt{}\Rm{}{\it{}subgroup\_K}\nwendquote} to fill in parts of the
light cone.

\nwenddocs{}\nwbegincode{91}\sublabel{NW3OF8UE-2KWnYy-1}\nwmargintag{{\nwtagstyle{}\subpageref{NW3OF8UE-2KWnYy-1}}}\moddef{Build future-past transition~{\nwtagstyle{}\subpageref{NW3OF8UE-2KWnYy-1}}}\endmoddef\Rm{}\nwstartdeflinemarkup\nwusesondefline{\\{NW3OF8UE-1p0Y9w-4}}\nwprevnextdefs{\relax}{NW3OF8UE-2KWnYy-2}\nwenddeflinemarkup
{\bf{}const} {\bf{}int} {\it{}curves} = 15, {\it{}nodes} = 40, {\it{}frames} = 8;\nwindexdefn{\nwixident{curves}}{curves}{NW3OF8UE-2KWnYy-1}
{\bf{}const} {\bf{}float} {\it{}exp\_scale} = 1.3, {\it{}node\_scale} = 4.0,\nwindexdefn{\nwixident{exp{\_}scale}}{exp:unscale}{NW3OF8UE-2KWnYy-1}
    {\it{}rad}[] = {\nwlbrace}1.0\begin{math}\div\end{math}5, 1.0\begin{math}\div\end{math}4, 1\begin{math}\div\end{math}3.5, 1.0\begin{math}\div\end{math}3, 1\begin{math}\div\end{math}2.5, 1.0\begin{math}\div\end{math}2, 1\begin{math}\div\end{math}1.5, 1, 1.5, 2, 2.5, 3, 3.5, 4, 4.5, 5{\nwrbrace};
{\bf{}char} {\it{}name}[] = {\tt{}"future-past-00.d"};\nwindexdefn{\nwixident{name}}{name}{NW3OF8UE-2KWnYy-1}
{\bf{}char}\begin{math}\ast\end{math} {\it{}S} = {\it{}name};\nwindexdefn{\nwixident{S}}{S}{NW3OF8UE-2KWnYy-1}
{\it{}ulim} = 8.5;
{\it{}vlim} = 8.5;

{\it{}metric} = {\it{}hyperbolic};
{\it{}subgroup} = {\it{}subgroup\_K};
{\it{}direct} = {\bf{}true};
{\it{}inversion} = {\bf{}true}; // to cut around unit circle.
\LA{}Initialise Clifford numbers~{\nwtagstyle{}\subpageref{NW3OF8UE-2xd863-1}}\RA{}
{\bf{}ex} {\it{}Fut} = {\it{}clifford\_moebius\_map}({\it{}dirac\_ONE}(), -{\it{}a} \begin{math}\ast\end{math} {\it{}e1},
                                   {\it{}a} \begin{math}\ast\end{math} {\it{}e1}, {\it{}dirac\_ONE}(), {\bf{}lst}({\it{}x}, {\it{}y}), {\it{}M});

\nwalsodefined{\\{NW3OF8UE-2KWnYy-2}}\nwused{\\{NW3OF8UE-1p0Y9w-4}}\nwidentdefs{\\{{\nwixident{curves}}{curves}}\\{{\nwixident{exp{\_}scale}}{exp:unscale}}\\{{\nwixident{name}}{name}}\\{{\nwixident{S}}{S}}}\nwidentuses{\\{{\nwixident{hyperbolic}}{hyperbolic}}\\{{\nwixident{metric}}{metric}}\\{{\nwixident{subgroup}}{subgroup}}\\{{\nwixident{subgroup{\_}K}}{subgroup:unK}}\\{{\nwixident{ulim}}{ulim}}}\nwindexuse{\nwixident{hyperbolic}}{hyperbolic}{NW3OF8UE-2KWnYy-1}\nwindexuse{\nwixident{metric}}{metric}{NW3OF8UE-2KWnYy-1}\nwindexuse{\nwixident{subgroup}}{subgroup}{NW3OF8UE-2KWnYy-1}\nwindexuse{\nwixident{subgroup{\_}K}}{subgroup:unK}{NW3OF8UE-2KWnYy-1}\nwindexuse{\nwixident{ulim}}{ulim}{NW3OF8UE-2KWnYy-1}\nwendcode{}\nwbegindocs{92}Then we proceed with iteration over the frames.

\nwenddocs{}\nwbegincode{93}\sublabel{NW3OF8UE-2KWnYy-2}\nwmargintag{{\nwtagstyle{}\subpageref{NW3OF8UE-2KWnYy-2}}}\moddef{Build future-past transition~{\nwtagstyle{}\subpageref{NW3OF8UE-2KWnYy-1}}}\plusendmoddef\Rm{}\nwstartdeflinemarkup\nwusesondefline{\\{NW3OF8UE-1p0Y9w-4}}\nwprevnextdefs{NW3OF8UE-2KWnYy-1}{\relax}\nwenddeflinemarkup
{\bf{}for} ({\bf{}int} {\it{}j} = 0; {\it{}j} \begin{math}<\end{math} {\it{}frames}; {\it{}j}\protect\PP) {\nwlbrace} // the power of transformation
        {\it{}sprintf}({\it{}S}, {\tt{}"future-past-
        {\it{}fileout}[0] = {\it{}fopen}({\it{}S}, {\tt{}"w"});
        {\bf{}for} ({\bf{}int} {\it{}k} = 0; {\it{}k} \begin{math}<\end{math} {\it{}curves}; {\it{}k}\protect\PP) {\nwlbrace} // the number of  curve
                {\it{}color\_grade} = {\it{}k}\begin{math}\div\end{math}{\it{}frames};
                {\it{}init\_coord}(0);
                \LA{}Iteration over a curve~{\nwtagstyle{}\subpageref{NW3OF8UE-3DeWAC-1}}\RA{}
                {\it{}close\_curve}(0);
        {\nwrbrace}
        {\it{}fclose}({\it{}fileout}[0]);
{\nwrbrace}
{\it{}cout} \begin{math}\ll\end{math} {\it{}endl};

\nwused{\\{NW3OF8UE-1p0Y9w-4}}\nwidentuses{\\{{\nwixident{close{\_}curve}}{close:uncurve}}\\{{\nwixident{curves}}{curves}}\\{{\nwixident{fileout}}{fileout}}\\{{\nwixident{init{\_}coord}}{init:uncoord}}\\{{\nwixident{S}}{S}}}\nwindexuse{\nwixident{close{\_}curve}}{close:uncurve}{NW3OF8UE-2KWnYy-2}\nwindexuse{\nwixident{curves}}{curves}{NW3OF8UE-2KWnYy-2}\nwindexuse{\nwixident{fileout}}{fileout}{NW3OF8UE-2KWnYy-2}\nwindexuse{\nwixident{init{\_}coord}}{init:uncoord}{NW3OF8UE-2KWnYy-2}\nwindexuse{\nwixident{S}}{S}{NW3OF8UE-2KWnYy-2}\nwendcode{}\nwbegindocs{94}A curve is created by rotation in hyperbolic metric and then a
M\"obius transformations  corresponding to the frame number is applied.
\nwenddocs{}\nwbegincode{95}\sublabel{NW3OF8UE-3DeWAC-1}\nwmargintag{{\nwtagstyle{}\subpageref{NW3OF8UE-3DeWAC-1}}}\moddef{Iteration over a curve~{\nwtagstyle{}\subpageref{NW3OF8UE-3DeWAC-1}}}\endmoddef\Rm{}\nwstartdeflinemarkup\nwusesondefline{\\{NW3OF8UE-2KWnYy-2}}\nwenddeflinemarkup
{\bf{}for} ({\bf{}int} {\it{}l} = -{\it{}nodes}\begin{math}\div\end{math}2; {\it{}l} \begin{math}\leq\end{math} {\it{}nodes}\begin{math}\div\end{math}2; {\it{}l}\protect\PP)  // the node number
    {\bf{}try} {\nwlbrace} // There is a chance of singularity!
        {\bf{}float} {\it{}angl} = (({\it{}j} \begin{math}>\end{math} 0) ? {\it{}exp}({\bf{}double}({\it{}j}\begin{math}\div\end{math}{\it{}exp\_scale}-3)) : 0);
        {\it{}res} = {\it{}Fut}.{\it{}subs}({\bf{}lst}( {\it{}a} \begin{math}\equiv\end{math} {\it{}angl}, {\it{}x} \begin{math}\equiv\end{math} {\it{}rad}[{\it{}k}]\begin{math}\ast\end{math}{\it{}cosh}({\it{}l}\begin{math}\div\end{math}{\it{}node\_scale}), {\it{}y} \begin{math}\equiv\end{math} {\it{}rad}[{\it{}k}]\begin{math}\ast\end{math}{\it{}sinh}({\it{}l}\begin{math}\div\end{math}{\it{}node\_scale})));
        {\it{}get\_components};
        {\it{}if\_in\_limits}(0);
    {\nwrbrace} {\bf{}catch}  ({\it{}exception} &{\it{}p}) {\nwlbrace}
            {\it{}catch\_handle}(0);
    {\nwrbrace}

\nwused{\\{NW3OF8UE-2KWnYy-2}}\nwidentuses{\\{{\nwixident{catch}}{catch}}\\{{\nwixident{catch{\_}handle}}{catch:unhandle}}\\{{\nwixident{exp{\_}scale}}{exp:unscale}}\\{{\nwixident{get{\_}components}}{get:uncomponents}}\\{{\nwixident{if{\_}in{\_}limits}}{if:unin:unlimits}}}\nwindexuse{\nwixident{catch}}{catch}{NW3OF8UE-3DeWAC-1}\nwindexuse{\nwixident{catch{\_}handle}}{catch:unhandle}{NW3OF8UE-3DeWAC-1}\nwindexuse{\nwixident{exp{\_}scale}}{exp:unscale}{NW3OF8UE-3DeWAC-1}\nwindexuse{\nwixident{get{\_}components}}{get:uncomponents}{NW3OF8UE-3DeWAC-1}\nwindexuse{\nwixident{if{\_}in{\_}limits}}{if:unin:unlimits}{NW3OF8UE-3DeWAC-1}\nwendcode{}\nwbegindocs{96}\nwdocspar
\subsection{Single Node Calculation}
\label{sec:single-node-calc}
This is a common portion of code for building orbits and transverses.
A single entry into \MetaPost\  file is calculated and
written. Calculations depend for three possible values of the
{\Tt{}\Rm{}{\it{}subgroup}\nwendquote}.  For each of them we create a special list {\Tt{}\Rm{}{\it{}a\_node}\nwendquote} of
substitutions for the already symbolically calculated {\Tt{}\Rm{}{\it{}Moebius}[][]\nwendquote}.

\nwenddocs{}\nwbegindocs{97}For subgroup \(A\) the points are distributed evenly on the unit circle.
\nwenddocs{}\nwbegincode{98}\sublabel{NW3OF8UE-4P0NfT-1}\nwmargintag{{\nwtagstyle{}\subpageref{NW3OF8UE-4P0NfT-1}}}\moddef{Generating one entry~{\nwtagstyle{}\subpageref{NW3OF8UE-4P0NfT-1}}}\endmoddef\Rm{}\nwstartdeflinemarkup\nwusesondefline{\\{NW3OF8UE-bYD0P-1}\\{NW3OF8UE-38hDUf-2}}\nwprevnextdefs{\relax}{NW3OF8UE-4P0NfT-2}\nwenddeflinemarkup
{\bf{}switch} ({\it{}subgroup}) {\nwlbrace}
{\bf{}case} {\it{}subgroup\_A}:
    {\it{}vval} = 1.0\begin{math}\ast\end{math}{\it{}vi}\begin{math}\div\end{math}({\it{}vilimits}[{\it{}subgroup}][{\it{}metric}]-1);
    {\bf{}if} ({\it{}metric} \begin{math}\equiv\end{math} {\it{}hyperbolic}) // we need a double set of value for negatives as well
        {\it{}vval} \begin{math}\ast\end{math}= 2;
    {\it{}a\_node} = {\bf{}lst}({\it{}t} \begin{math}\equiv\end{math} {\it{}f}, {\it{}x} \begin{math}\equiv\end{math} {\it{}cos}({\it{}Pi}\begin{math}\ast\end{math}{\it{}vval}), {\it{}y} \begin{math}\equiv\end{math} {\it{}sin}({\it{}Pi}\begin{math}\ast\end{math}{\it{}vval}));
    {\bf{}break};

\nwalsodefined{\\{NW3OF8UE-4P0NfT-2}\\{NW3OF8UE-4P0NfT-3}\\{NW3OF8UE-4P0NfT-4}}\nwused{\\{NW3OF8UE-bYD0P-1}\\{NW3OF8UE-38hDUf-2}}\nwidentuses{\\{{\nwixident{hyperbolic}}{hyperbolic}}\\{{\nwixident{metric}}{metric}}\\{{\nwixident{subgroup}}{subgroup}}\\{{\nwixident{subgroup{\_}A}}{subgroup:unA}}\\{{\nwixident{vilimits}}{vilimits}}}\nwindexuse{\nwixident{hyperbolic}}{hyperbolic}{NW3OF8UE-4P0NfT-1}\nwindexuse{\nwixident{metric}}{metric}{NW3OF8UE-4P0NfT-1}\nwindexuse{\nwixident{subgroup}}{subgroup}{NW3OF8UE-4P0NfT-1}\nwindexuse{\nwixident{subgroup{\_}A}}{subgroup:unA}{NW3OF8UE-4P0NfT-1}\nwindexuse{\nwixident{vilimits}}{vilimits}{NW3OF8UE-4P0NfT-1}\nwendcode{}\nwbegindocs{99}For the subgroups \(K\) and \(N\) points are distributed evenly on the vertical
axis.
\nwenddocs{}\nwbegincode{100}\sublabel{NW3OF8UE-4P0NfT-2}\nwmargintag{{\nwtagstyle{}\subpageref{NW3OF8UE-4P0NfT-2}}}\moddef{Generating one entry~{\nwtagstyle{}\subpageref{NW3OF8UE-4P0NfT-1}}}\plusendmoddef\Rm{}\nwstartdeflinemarkup\nwusesondefline{\\{NW3OF8UE-bYD0P-1}\\{NW3OF8UE-38hDUf-2}}\nwprevnextdefs{NW3OF8UE-4P0NfT-1}{NW3OF8UE-4P0NfT-3}\nwenddeflinemarkup
{\bf{}case} {\it{}subgroup\_K}:
    {\it{}vval} = {\it{}vpoints}[{\it{}metric}][{\it{}vi}];
    {\it{}a\_node} = {\bf{}lst}({\it{}t} \begin{math}\equiv\end{math} ({\it{}f} \begin{math}\ast\end{math} {\it{}Pi}), {\it{}x} \begin{math}\equiv\end{math} 0, {\it{}y} \begin{math}\equiv\end{math} {\it{}vval});
    {\bf{}break};

\nwused{\\{NW3OF8UE-bYD0P-1}\\{NW3OF8UE-38hDUf-2}}\nwidentuses{\\{{\nwixident{metric}}{metric}}\\{{\nwixident{subgroup{\_}K}}{subgroup:unK}}}\nwindexuse{\nwixident{metric}}{metric}{NW3OF8UE-4P0NfT-2}\nwindexuse{\nwixident{subgroup{\_}K}}{subgroup:unK}{NW3OF8UE-4P0NfT-2}\nwendcode{}\nwbegindocs{101} For the subgroup \(N\) we need  additional  points.
\nwenddocs{}\nwbegincode{102}\sublabel{NW3OF8UE-4P0NfT-3}\nwmargintag{{\nwtagstyle{}\subpageref{NW3OF8UE-4P0NfT-3}}}\moddef{Generating one entry~{\nwtagstyle{}\subpageref{NW3OF8UE-4P0NfT-1}}}\plusendmoddef\Rm{}\nwstartdeflinemarkup\nwusesondefline{\\{NW3OF8UE-bYD0P-1}\\{NW3OF8UE-38hDUf-2}}\nwprevnextdefs{NW3OF8UE-4P0NfT-2}{NW3OF8UE-4P0NfT-4}\nwenddeflinemarkup
{\bf{}case} {\it{}subgroup\_N}:
    {\bf{}if} ({\it{}metric} \begin{math}\equiv\end{math} {\it{}hyperbolic}) {\nwlbrace}// we need a double set of value for negatives as well
        {\it{}vval} = ( (({\it{}vi} - {\it{}vilimits}[{\it{}subgroup}][{\it{}metric}]\begin{math}\div\end{math}2) \begin{math}<\end{math} 0 ) ? -1 : 1)
            \begin{math}\ast\end{math}{\it{}vpoints}[{\it{}metric}][{\it{}abs}({\it{}vi} - {\it{}vilimits}[{\it{}subgroup}][{\it{}metric}]\begin{math}\div\end{math}2)];
    {\nwrbrace} {\bf{}else}
        {\it{}vval} = {\it{}vpoints}[{\it{}metric}][{\it{}vi}];
    {\it{}a\_node} = {\bf{}lst}({\it{}t} \begin{math}\equiv\end{math} {\it{}f}, {\it{}x} \begin{math}\equiv\end{math} 0, {\it{}y} \begin{math}\equiv\end{math} {\it{}vval});
    {\bf{}break};
{\nwrbrace}

\nwused{\\{NW3OF8UE-bYD0P-1}\\{NW3OF8UE-38hDUf-2}}\nwidentuses{\\{{\nwixident{hyperbolic}}{hyperbolic}}\\{{\nwixident{metric}}{metric}}\\{{\nwixident{subgroup}}{subgroup}}\\{{\nwixident{subgroup{\_}N}}{subgroup:unN}}\\{{\nwixident{vilimits}}{vilimits}}}\nwindexuse{\nwixident{hyperbolic}}{hyperbolic}{NW3OF8UE-4P0NfT-3}\nwindexuse{\nwixident{metric}}{metric}{NW3OF8UE-4P0NfT-3}\nwindexuse{\nwixident{subgroup}}{subgroup}{NW3OF8UE-4P0NfT-3}\nwindexuse{\nwixident{subgroup{\_}N}}{subgroup:unN}{NW3OF8UE-4P0NfT-3}\nwindexuse{\nwixident{vilimits}}{vilimits}{NW3OF8UE-4P0NfT-3}\nwendcode{}\nwbegindocs{103}And now using values stored above in {\Tt{}\Rm{}{\it{}a\_node}\nwendquote} we do the actual
calculation through substitution and write the node into the \MetaPost\ file.
\nwenddocs{}\nwbegincode{104}\sublabel{NW3OF8UE-4P0NfT-4}\nwmargintag{{\nwtagstyle{}\subpageref{NW3OF8UE-4P0NfT-4}}}\moddef{Generating one entry~{\nwtagstyle{}\subpageref{NW3OF8UE-4P0NfT-1}}}\plusendmoddef\Rm{}\nwstartdeflinemarkup\nwusesondefline{\\{NW3OF8UE-bYD0P-1}\\{NW3OF8UE-38hDUf-2}}\nwprevnextdefs{NW3OF8UE-4P0NfT-3}{\relax}\nwenddeflinemarkup
{\bf{}try}{\nwlbrace}
    {\it{}res} = {\it{}Moebius}[{\it{}subgroup}][0].{\it{}subs}({\it{}a\_node});
    {\it{}transverse\_dir}(0);
    {\it{}get\_components};
    {\it{}if\_in\_limits}(0);
{\nwrbrace} {\bf{}catch}  ({\it{}exception} &{\it{}p}) {\nwlbrace}\nwindexdefn{\nwixident{catch}}{catch}{NW3OF8UE-4P0NfT-4}
    {\it{}catch\_handle}(0);
{\nwrbrace}

\nwused{\\{NW3OF8UE-bYD0P-1}\\{NW3OF8UE-38hDUf-2}}\nwidentdefs{\\{{\nwixident{catch}}{catch}}}\nwidentuses{\\{{\nwixident{catch{\_}handle}}{catch:unhandle}}\\{{\nwixident{get{\_}components}}{get:uncomponents}}\\{{\nwixident{if{\_}in{\_}limits}}{if:unin:unlimits}}\\{{\nwixident{subgroup}}{subgroup}}\\{{\nwixident{transverse{\_}dir}}{transverse:undir}}}\nwindexuse{\nwixident{catch{\_}handle}}{catch:unhandle}{NW3OF8UE-4P0NfT-4}\nwindexuse{\nwixident{get{\_}components}}{get:uncomponents}{NW3OF8UE-4P0NfT-4}\nwindexuse{\nwixident{if{\_}in{\_}limits}}{if:unin:unlimits}{NW3OF8UE-4P0NfT-4}\nwindexuse{\nwixident{subgroup}}{subgroup}{NW3OF8UE-4P0NfT-4}\nwindexuse{\nwixident{transverse{\_}dir}}{transverse:undir}{NW3OF8UE-4P0NfT-4}\nwendcode{}\nwbegindocs{105}We define the tangents to the transverse lines here.
\nwenddocs{}\nwbegincode{106}\sublabel{NW3OF8UE-4LL0eE-1}\nwmargintag{{\nwtagstyle{}\subpageref{NW3OF8UE-4LL0eE-1}}}\moddef{Define transverse directions~{\nwtagstyle{}\subpageref{NW3OF8UE-4LL0eE-1}}}\endmoddef\Rm{}\nwstartdeflinemarkup\nwusesondefline{\\{NW3OF8UE-38hDUf-1}}\nwprevnextdefs{\relax}{NW3OF8UE-4LL0eE-2}\nwenddeflinemarkup
{\bf{}switch} ({\it{}subgroup}) {\nwlbrace}
{\bf{}case} {\it{}subgroup\_A}:
    {\it{}a\_trans} = {\bf{}lst}({\it{}tr\_u} \begin{math}\equiv\end{math} -{\it{}y}, {\it{}tr\_v} \begin{math}\equiv\end{math} {\it{}x});
    {\bf{}break};
{\bf{}case} {\it{}subgroup\_N}:
{\bf{}case} {\it{}subgroup\_K}:
    {\it{}a\_trans} = {\bf{}lst}({\it{}tr\_u} \begin{math}\equiv\end{math} 0, {\it{}tr\_v} \begin{math}\equiv\end{math} 1);
    {\bf{}break};
{\nwrbrace}

\nwalsodefined{\\{NW3OF8UE-4LL0eE-2}}\nwused{\\{NW3OF8UE-38hDUf-1}}\nwidentuses{\\{{\nwixident{subgroup}}{subgroup}}\\{{\nwixident{subgroup{\_}A}}{subgroup:unA}}\\{{\nwixident{subgroup{\_}K}}{subgroup:unK}}\\{{\nwixident{subgroup{\_}N}}{subgroup:unN}}}\nwindexuse{\nwixident{subgroup}}{subgroup}{NW3OF8UE-4LL0eE-1}\nwindexuse{\nwixident{subgroup{\_}A}}{subgroup:unA}{NW3OF8UE-4LL0eE-1}\nwindexuse{\nwixident{subgroup{\_}K}}{subgroup:unK}{NW3OF8UE-4LL0eE-1}\nwindexuse{\nwixident{subgroup{\_}N}}{subgroup:unN}{NW3OF8UE-4LL0eE-1}\nwendcode{}\nwbegindocs{107}And here again comes evaluation through substitution.
\nwenddocs{}\nwbegincode{108}\sublabel{NW3OF8UE-4LL0eE-2}\nwmargintag{{\nwtagstyle{}\subpageref{NW3OF8UE-4LL0eE-2}}}\moddef{Define transverse directions~{\nwtagstyle{}\subpageref{NW3OF8UE-4LL0eE-1}}}\plusendmoddef\Rm{}\nwstartdeflinemarkup\nwusesondefline{\\{NW3OF8UE-38hDUf-1}}\nwprevnextdefs{NW3OF8UE-4LL0eE-1}{\relax}\nwenddeflinemarkup
{\bf{}for} ({\bf{}int} {\it{}cal}=0; {\it{}cal} \begin{math}<\end{math} 3; {\it{}cal}\protect\PP)
 {\it{}trans\_dir\_sub}[{\it{}cal}] = {\bf{}matrix}(2, 1,
        {\bf{}lst}({\it{}trans\_dir}[{\it{}subgroup}][{\it{}cal}].{\it{}op}(0).{\it{}subs}({\it{}a\_trans}).{\it{}normal}(),
         {\it{}trans\_dir}[{\it{}subgroup}][{\it{}cal}].{\it{}op}(1).{\it{}subs}({\it{}a\_trans}).{\it{}normal}()));

\nwused{\\{NW3OF8UE-38hDUf-1}}\nwidentuses{\\{{\nwixident{subgroup}}{subgroup}}}\nwindexuse{\nwixident{subgroup}}{subgroup}{NW3OF8UE-4LL0eE-2}\nwendcode{}\nwbegindocs{109}\nwdocspar
\subsection{Cayley Transforms of Images}
\label{sec:cayl-transf-imag}
We will need a calculation parameters of parabola into the Cayley
transform images. This checks the
statement~\cite[Lemma~\ref{E-le:parabolic-disk}]{Kisil05a}. The
calculation is done by linear equation solver {\Tt{}\Rm{}{\it{}lsolve}()\nwendquote} from \GiNaC.
\nwenddocs{}\nwbegincode{110}\sublabel{NW3OF8UE-4TccJC-9}\nwmargintag{{\nwtagstyle{}\subpageref{NW3OF8UE-4TccJC-9}}}\moddef{Definitions~{\nwtagstyle{}\subpageref{NW3OF8UE-4TccJC-1}}}\plusendmoddef\Rm{}\nwstartdeflinemarkup\nwusesondefline{\\{NW3OF8UE-1p0Y9w-1}}\nwprevnextdefs{NW3OF8UE-4TccJC-8}{NW3OF8UE-4TccJC-A}\nwenddeflinemarkup
{\bf{}\char35{}define}{\tt{} calc\_par\_focal(X) if (direct && (metric == parabolic)  \begin{math}\backslash\end{math}}\nwindexdefn{\nwixident{calc{\_}par{\_}focal}}{calc:unpar:unfocal}{NW3OF8UE-4TccJC-9}
                              \begin{math}\wedge\end{math} ({\it{}subgroup} \begin{math}\neq\end{math} {\it{}subgroup\_K})) {\nwlbrace}   \begin{math}\backslash\end{math}
        {\it{}up}[2][{\it{}X}] = {\it{}u\_res};                                \begin{math}\backslash\end{math}
        {\it{}vp}[2][{\it{}X}] = {\it{}v\_res};                                        \begin{math}\backslash\end{math}
        {\bf{}if} ({\it{}j} \begin{math}\equiv\end{math} 1) {\nwlbrace}                                            \begin{math}\backslash\end{math}
            {\bf{}lst} {\it{}eqns}, {\it{}vars}  ;                                \begin{math}\backslash\end{math}
            {\it{}vars} = {\it{}a}, {\it{}b}, {\it{}c};                                  \begin{math}\backslash\end{math}
            {\it{}eqns} = {\it{}a}\begin{math}\ast\end{math}{\it{}pow}({\it{}up}[0][{\it{}X}], 2) + {\it{}b}\begin{math}\ast\end{math}{\it{}up}[0][{\it{}X}] + {\it{}c} \begin{math}\equiv\end{math} {\it{}vp}[0][{\it{}X}], \begin{math}\backslash\end{math}
                    {\it{}a}\begin{math}\ast\end{math}{\it{}pow}({\it{}up}[1][{\it{}X}], 2) + {\it{}b}\begin{math}\ast\end{math}{\it{}up}[1][{\it{}X}] + {\it{}c} \begin{math}\equiv\end{math} {\it{}vp}[1][{\it{}X}],    \begin{math}\backslash\end{math}
                    {\it{}a}\begin{math}\ast\end{math}{\it{}pow}({\it{}up}[2][{\it{}X}], 2) + {\it{}b}\begin{math}\ast\end{math}{\it{}up}[2][{\it{}X}] + {\it{}c} \begin{math}\equiv\end{math} {\it{}vp}[2][{\it{}X}];    \begin{math}\backslash\end{math}
            {\it{}soln}[{\it{}X}] = {\it{}ex\_to}\begin{math}<\end{math}{\bf{}lst}\begin{math}>\end{math}({\it{}lsolve}({\it{}eqns}, {\it{}vars}));                   \begin{math}\backslash\end{math}
        {\nwrbrace}                                                               \begin{math}\backslash\end{math}
{\commopen}After calculation is made we store previous values for the next round.{\commclose}\begin{math}\backslash\end{math}
        {\it{}up}[0][{\it{}X}] = {\it{}up}[1][{\it{}X}];                                            \begin{math}\backslash\end{math}
        {\it{}vp}[0][{\it{}X}] = {\it{}vp}[1][{\it{}X}];                                            \begin{math}\backslash\end{math}
        {\it{}up}[1][{\it{}X}] = {\it{}up}[2][{\it{}X}];                                            \begin{math}\backslash\end{math}
        {\it{}vp}[1][{\it{}X}] = {\it{}vp}[2][{\it{}X}];                                            \begin{math}\backslash\end{math}
    {\nwrbrace}

\nwused{\\{NW3OF8UE-1p0Y9w-1}}\nwidentdefs{\\{{\nwixident{calc{\_}par{\_}focal}}{calc:unpar:unfocal}}}\nwidentuses{\\{{\nwixident{metric}}{metric}}\\{{\nwixident{parabolic}}{parabolic}}\\{{\nwixident{subgroup}}{subgroup}}\\{{\nwixident{subgroup{\_}K}}{subgroup:unK}}}\nwindexuse{\nwixident{metric}}{metric}{NW3OF8UE-4TccJC-9}\nwindexuse{\nwixident{parabolic}}{parabolic}{NW3OF8UE-4TccJC-9}\nwindexuse{\nwixident{subgroup}}{subgroup}{NW3OF8UE-4TccJC-9}\nwindexuse{\nwixident{subgroup{\_}K}}{subgroup:unK}{NW3OF8UE-4TccJC-9}\nwendcode{}\nwbegindocs{111}\nwdocspar
We produce two versions of the Cayley transforms for each node of the orbit or
transverses lines. This done by simple substitution of {\Tt{}\Rm{}{\it{}a\_node}\nwendquote} into
{\Tt{}\Rm{}{\it{}Moebius}[][3,4]\nwendquote} symbolically calculated in
subsection~\ref{sec:mobi-transf-1}.
\nwenddocs{}\nwbegincode{112}\sublabel{NW3OF8UE-D1caK-1}\nwmargintag{{\nwtagstyle{}\subpageref{NW3OF8UE-D1caK-1}}}\moddef{Producing Cayley transform of the orbit~{\nwtagstyle{}\subpageref{NW3OF8UE-D1caK-1}}}\endmoddef\Rm{}\nwstartdeflinemarkup\nwusesondefline{\\{NW3OF8UE-bYD0P-1}\\{NW3OF8UE-38hDUf-2}}\nwprevnextdefs{\relax}{NW3OF8UE-D1caK-2}\nwenddeflinemarkup
{\it{}cayley} = {\bf{}true};
{\bf{}try} {\nwlbrace} // There is a chance of singularity!
        {\it{}res} = {\it{}Moebius}[{\it{}subgroup}][3].{\it{}subs}({\it{}a\_node});
    {\it{}transverse\_dir}(1);
        {\it{}get\_components};
        {\it{}if\_in\_limits}(1);
        {\it{}calc\_par\_focal}(0);
{\nwrbrace} {\bf{}catch}  ({\it{}exception} &{\it{}p}) {\nwlbrace}\nwindexdefn{\nwixident{catch}}{catch}{NW3OF8UE-D1caK-1}
        {\it{}catch\_handle}(1);
{\nwrbrace}
\nwalsodefined{\\{NW3OF8UE-D1caK-2}}\nwused{\\{NW3OF8UE-bYD0P-1}\\{NW3OF8UE-38hDUf-2}}\nwidentdefs{\\{{\nwixident{catch}}{catch}}}\nwidentuses{\\{{\nwixident{calc{\_}par{\_}focal}}{calc:unpar:unfocal}}\\{{\nwixident{catch{\_}handle}}{catch:unhandle}}\\{{\nwixident{get{\_}components}}{get:uncomponents}}\\{{\nwixident{if{\_}in{\_}limits}}{if:unin:unlimits}}\\{{\nwixident{subgroup}}{subgroup}}\\{{\nwixident{transverse{\_}dir}}{transverse:undir}}}\nwindexuse{\nwixident{calc{\_}par{\_}focal}}{calc:unpar:unfocal}{NW3OF8UE-D1caK-1}\nwindexuse{\nwixident{catch{\_}handle}}{catch:unhandle}{NW3OF8UE-D1caK-1}\nwindexuse{\nwixident{get{\_}components}}{get:uncomponents}{NW3OF8UE-D1caK-1}\nwindexuse{\nwixident{if{\_}in{\_}limits}}{if:unin:unlimits}{NW3OF8UE-D1caK-1}\nwindexuse{\nwixident{subgroup}}{subgroup}{NW3OF8UE-D1caK-1}\nwindexuse{\nwixident{transverse{\_}dir}}{transverse:undir}{NW3OF8UE-D1caK-1}\nwendcode{}\nwbegindocs{113}For second type of the Cayley transforms  we perform  an extra run
similar to the above.
\nwenddocs{}\nwbegincode{114}\sublabel{NW3OF8UE-D1caK-2}\nwmargintag{{\nwtagstyle{}\subpageref{NW3OF8UE-D1caK-2}}}\moddef{Producing Cayley transform of the orbit~{\nwtagstyle{}\subpageref{NW3OF8UE-D1caK-1}}}\plusendmoddef\Rm{}\nwstartdeflinemarkup\nwusesondefline{\\{NW3OF8UE-bYD0P-1}\\{NW3OF8UE-38hDUf-2}}\nwprevnextdefs{NW3OF8UE-D1caK-1}{\relax}\nwenddeflinemarkup
{\bf{}try} {\nwlbrace}
        {\it{}res} = {\it{}Moebius}[{\it{}subgroup}][4].{\it{}subs}({\it{}a\_node});
    {\it{}transverse\_dir}(2);
        {\it{}get\_components};
        {\it{}if\_in\_limits}(2);
        {\it{}calc\_par\_focal}(1);
{\nwrbrace} {\bf{}catch}  ({\it{}exception} &{\it{}p}) {\nwlbrace}\nwindexdefn{\nwixident{catch}}{catch}{NW3OF8UE-D1caK-2}
        {\it{}catch\_handle}(2);
{\nwrbrace}
{\it{}cayley} = {\bf{}false};

\nwused{\\{NW3OF8UE-bYD0P-1}\\{NW3OF8UE-38hDUf-2}}\nwidentdefs{\\{{\nwixident{catch}}{catch}}}\nwidentuses{\\{{\nwixident{calc{\_}par{\_}focal}}{calc:unpar:unfocal}}\\{{\nwixident{catch{\_}handle}}{catch:unhandle}}\\{{\nwixident{get{\_}components}}{get:uncomponents}}\\{{\nwixident{if{\_}in{\_}limits}}{if:unin:unlimits}}\\{{\nwixident{subgroup}}{subgroup}}\\{{\nwixident{transverse{\_}dir}}{transverse:undir}}}\nwindexuse{\nwixident{calc{\_}par{\_}focal}}{calc:unpar:unfocal}{NW3OF8UE-D1caK-2}\nwindexuse{\nwixident{catch{\_}handle}}{catch:unhandle}{NW3OF8UE-D1caK-2}\nwindexuse{\nwixident{get{\_}components}}{get:uncomponents}{NW3OF8UE-D1caK-2}\nwindexuse{\nwixident{if{\_}in{\_}limits}}{if:unin:unlimits}{NW3OF8UE-D1caK-2}\nwindexuse{\nwixident{subgroup}}{subgroup}{NW3OF8UE-D1caK-2}\nwindexuse{\nwixident{transverse{\_}dir}}{transverse:undir}{NW3OF8UE-D1caK-2}\nwendcode{}\nwbegindocs{115}\nwdocspar
\subsection{Numeric Check of Formulae}
\label{sec:numer-check-form}
Here is a numeric check of few formulas in the paper about radius
and focal length of sections. We calculate focal properties for three
types of orbits (circles, parabolas and hyperbolas) of the
{\Tt{}\Rm{}{\it{}subgroup\_K}\nwendquote}.
\nwenddocs{}\nwbegincode{116}\sublabel{NW3OF8UE-47Sd20-1}\nwmargintag{{\nwtagstyle{}\subpageref{NW3OF8UE-47Sd20-1}}}\moddef{Check formulas in the paper~{\nwtagstyle{}\subpageref{NW3OF8UE-47Sd20-1}}}\endmoddef\Rm{}\nwstartdeflinemarkup\nwusesondefline{\\{NW3OF8UE-bYD0P-1}}\nwprevnextdefs{\relax}{NW3OF8UE-47Sd20-2}\nwenddeflinemarkup
{\bf{}if} (({\it{}j} \begin{math}\neq\end{math} -{\it{}fsteps}[{\it{}subgroup}][{\it{}metric}]) \begin{math}\wedge\end{math} ({\it{}j} \begin{math}\neq\end{math} {\it{}fsteps}[{\it{}subgroup}][{\it{}metric}]) // End points are weird!
    \begin{math}\wedge\end{math} ({\it{}subgroup} \begin{math}\equiv\end{math} {\it{}subgroup\_K}) \begin{math}\wedge\end{math} ({\it{}vval} \begin{math}\neq\end{math} 0))  {\nwlbrace}// only for that values
    {\bf{}try} {\nwlbrace}
        {\bf{}switch}  ({\it{}metric}) {\nwlbrace}  // depends from the type of metric

\nwalsodefined{\\{NW3OF8UE-47Sd20-2}\\{NW3OF8UE-47Sd20-3}\\{NW3OF8UE-47Sd20-4}\\{NW3OF8UE-47Sd20-5}}\nwused{\\{NW3OF8UE-bYD0P-1}}\nwidentuses{\\{{\nwixident{metric}}{metric}}\\{{\nwixident{subgroup}}{subgroup}}\\{{\nwixident{subgroup{\_}K}}{subgroup:unK}}}\nwindexuse{\nwixident{metric}}{metric}{NW3OF8UE-47Sd20-1}\nwindexuse{\nwixident{subgroup}}{subgroup}{NW3OF8UE-47Sd20-1}\nwindexuse{\nwixident{subgroup{\_}K}}{subgroup:unK}{NW3OF8UE-47Sd20-1}\nwendcode{}\nwbegindocs{117}The values are calculated for each node on the orbit as follows,
see~\cite[Lemma~\ref{E-le:k-action}]{Kisil05a}.
    For {\Tt{}\Rm{}{\it{}elliptic}\nwendquote} orbits of {\Tt{}\Rm{}{\it{}sungroup\_K}\nwendquote}: a circle with the centre
     at \((0, (v+v^{-1})/2)\)  and the radius \((v-v^{-1})/2\).\\
\nwenddocs{}\nwbegincode{118}\sublabel{NW3OF8UE-47Sd20-2}\nwmargintag{{\nwtagstyle{}\subpageref{NW3OF8UE-47Sd20-2}}}\moddef{Check formulas in the paper~{\nwtagstyle{}\subpageref{NW3OF8UE-47Sd20-1}}}\plusendmoddef\Rm{}\nwstartdeflinemarkup\nwusesondefline{\\{NW3OF8UE-bYD0P-1}}\nwprevnextdefs{NW3OF8UE-47Sd20-1}{NW3OF8UE-47Sd20-3}\nwenddeflinemarkup
        {\bf{}case} {\it{}elliptic}:
            {\it{}focal\_f}[1] = {\it{}ex\_to}\begin{math}<\end{math}{\bf{}numeric}\begin{math}>\end{math}({\it{}pow}({\it{}u\_res}\begin{math}\ast\end{math}{\it{}u\_res}
                            + {\it{}pow}({\it{}v\_res}-({\it{}vval}+1\begin{math}\div\end{math}{\it{}vval})\begin{math}\div\end{math}2, 2) , 0.5)).{\it{}to\_double}();
            {\bf{}break};
\nwused{\\{NW3OF8UE-bYD0P-1}}\nwidentuses{\\{{\nwixident{elliptic}}{elliptic}}\\{{\nwixident{numeric}}{numeric}}}\nwindexuse{\nwixident{elliptic}}{elliptic}{NW3OF8UE-47Sd20-2}\nwindexuse{\nwixident{numeric}}{numeric}{NW3OF8UE-47Sd20-2}\nwendcode{}\nwbegindocs{119}   For {\Tt{}\Rm{}{\it{}parabolic}\nwendquote} orbits of {\Tt{}\Rm{}{\it{}sungroup\_K}\nwendquote}: a parabola with the focus at
    \((0,(v+v^{-1})/2)\) and focal length \(v^{-1}/2\).
\nwenddocs{}\nwbegincode{120}\sublabel{NW3OF8UE-47Sd20-3}\nwmargintag{{\nwtagstyle{}\subpageref{NW3OF8UE-47Sd20-3}}}\moddef{Check formulas in the paper~{\nwtagstyle{}\subpageref{NW3OF8UE-47Sd20-1}}}\plusendmoddef\Rm{}\nwstartdeflinemarkup\nwusesondefline{\\{NW3OF8UE-bYD0P-1}}\nwprevnextdefs{NW3OF8UE-47Sd20-2}{NW3OF8UE-47Sd20-4}\nwenddeflinemarkup
        {\bf{}case} {\it{}parabolic}:
            {\it{}focal\_f}[1] = {\it{}ex\_to}\begin{math}<\end{math}{\bf{}numeric}\begin{math}>\end{math}({\it{}pow}({\it{}u\_res}\begin{math}\ast\end{math}{\it{}u\_res}
                            + {\it{}pow}({\it{}v\_res}-({\it{}vval}+1\begin{math}\div\end{math}{\it{}vval}\begin{math}\div\end{math}4), 2) , 0.5)-{\it{}v\_res}).{\it{}to\_double}();
            {\bf{}break};
\nwused{\\{NW3OF8UE-bYD0P-1}}\nwidentuses{\\{{\nwixident{numeric}}{numeric}}\\{{\nwixident{parabolic}}{parabolic}}}\nwindexuse{\nwixident{numeric}}{numeric}{NW3OF8UE-47Sd20-3}\nwindexuse{\nwixident{parabolic}}{parabolic}{NW3OF8UE-47Sd20-3}\nwendcode{}\nwbegindocs{121}  For {\Tt{}\Rm{}{\it{}hyperbolic}\nwendquote} orbits of {\Tt{}\Rm{}{\it{}sungroup\_K}\nwendquote}: a hyperbola with the upper focus
    located at \((0,f)\) with:
    \begin{displaymath}
    f=\left\{\begin{array}{ll}
      p-\sqrt{\frac{p^2}{2}-1}, &\textrm{ for } 0<v<1; \textrm{
        and }\\
      p+\sqrt{\frac{p^2}{2}-1}, & \textrm{ for } v\geq 1.
    \end{array}\right.
  \end{displaymath}
  and has the focal distance between focuses \(2p\).

\nwenddocs{}\nwbegincode{122}\sublabel{NW3OF8UE-47Sd20-4}\nwmargintag{{\nwtagstyle{}\subpageref{NW3OF8UE-47Sd20-4}}}\moddef{Check formulas in the paper~{\nwtagstyle{}\subpageref{NW3OF8UE-47Sd20-1}}}\plusendmoddef\Rm{}\nwstartdeflinemarkup\nwusesondefline{\\{NW3OF8UE-bYD0P-1}}\nwprevnextdefs{NW3OF8UE-47Sd20-3}{NW3OF8UE-47Sd20-5}\nwenddeflinemarkup
        {\bf{}case} {\it{}hyperbolic}:
            {\it{}p} = ({\it{}vval}\begin{math}\ast\end{math}{\it{}vval}+1)\begin{math}\div\end{math}{\it{}vval}\begin{math}\div\end{math}{\it{}pow}(2,0.5);
            {\it{}focal} = (({\it{}vval}\begin{math}<\end{math}1) ? {\it{}p} -{\it{}pow}({\it{}p}\begin{math}\ast\end{math}{\it{}p}\begin{math}\div\end{math}2-1,0.5) : {\it{}p}+{\it{}pow}({\it{}p}\begin{math}\ast\end{math}{\it{}p}\begin{math}\div\end{math}2-1,0.5));
            {\it{}focal\_f}[1] = {\it{}ex\_to}\begin{math}<\end{math}{\bf{}numeric}\begin{math}>\end{math}({\it{}pow}({\it{}u\_res}\begin{math}\ast\end{math}{\it{}u\_res} + {\it{}pow}({\it{}v\_res}-{\it{}focal}, 2) , 0.5)
                            -{\it{}pow}({\it{}u\_res}\begin{math}\ast\end{math}{\it{}u\_res} + {\it{}pow}({\it{}v\_res}-{\it{}focal}+2\begin{math}\ast\end{math}{\it{}p}, 2) , 0.5)).{\it{}to\_double}();
            {\bf{}break};
        {\nwrbrace}

\nwused{\\{NW3OF8UE-bYD0P-1}}\nwidentuses{\\{{\nwixident{hyperbolic}}{hyperbolic}}\\{{\nwixident{numeric}}{numeric}}}\nwindexuse{\nwixident{hyperbolic}}{hyperbolic}{NW3OF8UE-47Sd20-4}\nwindexuse{\nwixident{numeric}}{numeric}{NW3OF8UE-47Sd20-4}\nwendcode{}\nwbegindocs{123}If the obtained value is reasonably close to the previous one then
{\Tt{}\Rm{}=\nwendquote} sign is printed to {\Tt{}\Rm{}{\it{}stdout}\nwendquote}, otherwise the new value is
printed. This produce lines similar to the following:
\begin{webcode}Distance to center is: 3.938======================\end{webcode}
\begin{rem}
  Note that all check are passed smoothly (see
  Appendix~\ref{sec:append-text-outp}), however in the {\Tt{}\Rm{}{\it{}hyperbolic}\nwendquote}
  case there is ``V'' shape of switch from positive values to negative
  and back (with the same absolute value) like this:
  \begin{webcode}Difference to foci is: 2.000====== -2.000============== 2.000======\end{webcode}
  This demonstrates the
  non-invariance of the upper half plane in the {\Tt{}\Rm{}{\it{}hyperbolic}\nwendquote} case as
  explained in \cite[\S~\ref{E-sec:invar-upper-half}]{Kisil05a}.
\end{rem}
\nwenddocs{}\nwbegincode{124}\sublabel{NW3OF8UE-47Sd20-5}\nwmargintag{{\nwtagstyle{}\subpageref{NW3OF8UE-47Sd20-5}}}\moddef{Check formulas in the paper~{\nwtagstyle{}\subpageref{NW3OF8UE-47Sd20-1}}}\plusendmoddef\Rm{}\nwstartdeflinemarkup\nwusesondefline{\\{NW3OF8UE-bYD0P-1}}\nwprevnextdefs{NW3OF8UE-47Sd20-4}{\relax}\nwenddeflinemarkup
        {\bf{}if} (({\it{}abs}({\it{}focal\_f}[1] - {\it{}focal\_f}[0]) \begin{math}<\end{math} 0.001)
            \begin{math}\vee\end{math}  ({\it{}abs}({\it{}focal\_f}[1] - {\it{}focal\_f}[0]) \begin{math}<\end{math} 0.001\begin{math}\ast\end{math}({\it{}abs}({\it{}focal\_f}[1])+{\it{}abs}({\it{}focal\_f}[0]))))
            {\it{}cout} \begin{math}\ll\end{math} {\tt{}"="};
        {\bf{}else}
            {\it{}printf}({\tt{}" 
        {\it{}focal\_f}[0] = {\it{}focal\_f}[1];
    {\nwrbrace} {\bf{}catch}  ({\it{}exception} &{\it{}p}) {\nwlbrace}
        {\it{}cerr} \begin{math}\ll\end{math} {\tt{}"*** Got problem in formulas: "} \begin{math}\ll\end{math} {\it{}p}.{\it{}what}() \begin{math}\ll\end{math} {\it{}endl};
    {\nwrbrace}
{\nwrbrace}

\nwused{\\{NW3OF8UE-bYD0P-1}}\nwidentuses{\\{{\nwixident{catch}}{catch}}}\nwindexuse{\nwixident{catch}}{catch}{NW3OF8UE-47Sd20-5}\nwendcode{}\nwbegindocs{125}The last check we make is about some properties of Cayley transform
in {\Tt{}\Rm{}{\it{}parabolic}\nwendquote} case. All parameters of parabolic orbits were
calculated in Subsection~\ref{sec:cayl-transf-imag}, now we check
properties listed in~\cite[Lemma~\ref{E-le:parabolic-disk}]{Kisil05a}.
\nwenddocs{}\nwbegincode{126}\sublabel{NW3OF8UE-4TccJC-A}\nwmargintag{{\nwtagstyle{}\subpageref{NW3OF8UE-4TccJC-A}}}\moddef{Definitions~{\nwtagstyle{}\subpageref{NW3OF8UE-4TccJC-1}}}\plusendmoddef\Rm{}\nwstartdeflinemarkup\nwusesondefline{\\{NW3OF8UE-1p0Y9w-1}}\nwprevnextdefs{NW3OF8UE-4TccJC-9}{\relax}\nwenddeflinemarkup
{\bf{}\char35{}define}{\tt{} output\_focal(X) ex\_to\begin{math}<\end{math}numeric\begin{math}>\end{math}(focal\_u.subs(soln[X]).evalf()).to\_double(), \begin{math}\backslash\end{math}}\nwindexdefn{\nwixident{output{\_}focal}}{output:unfocal}{NW3OF8UE-4TccJC-A}
    {\it{}ex\_to}\begin{math}<\end{math}{\bf{}numeric}\begin{math}>\end{math}({\it{}focal\_v}.{\it{}subs}({\it{}soln}[{\it{}X}]).{\it{}evalf}()).{\it{}to\_double}(), \begin{math}\backslash\end{math}
    {\it{}ex\_to}\begin{math}<\end{math}{\bf{}numeric}\begin{math}>\end{math}({\it{}focal\_l}.{\it{}subs}({\it{}soln}[{\it{}X}]).{\it{}evalf}()).{\it{}to\_double}()

\nwused{\\{NW3OF8UE-1p0Y9w-1}}\nwidentdefs{\\{{\nwixident{output{\_}focal}}{output:unfocal}}}\nwidentuses{\\{{\nwixident{numeric}}{numeric}}}\nwindexuse{\nwixident{numeric}}{numeric}{NW3OF8UE-4TccJC-A}\nwendcode{}\nwbegindocs{127}Here is expressions for focal length {\Tt{}\Rm{}{\it{}focal\_l}\nwendquote} and vertex
({\Tt{}\Rm{}{\it{}focal\_u}\nwendquote}, {\Tt{}\Rm{}{\it{}focal\_v}\nwendquote}) of a parabola given by its equation \(v=
au^2+bu+c\).
\nwenddocs{}\nwbegincode{128}\sublabel{NW3OF8UE-2gOisp-1}\nwmargintag{{\nwtagstyle{}\subpageref{NW3OF8UE-2gOisp-1}}}\moddef{Parabola parameters~{\nwtagstyle{}\subpageref{NW3OF8UE-2gOisp-1}}}\endmoddef\Rm{}\nwstartdeflinemarkup\nwusesondefline{\\{NW3OF8UE-1p0Y9w-1}}\nwenddeflinemarkup
    {\it{}focal\_l} = 1\begin{math}\div\end{math}(4\begin{math}\ast\end{math}{\it{}a}); // focal length
    {\it{}focal\_u} = {\it{}b}\begin{math}\div\end{math}(2\begin{math}\ast\end{math}{\it{}a}); // \(u\) comp of focus
    {\it{}focal\_v} = {\it{}c}-{\it{}pow}({\it{}b}\begin{math}\div\end{math}(2\begin{math}\ast\end{math}{\it{}a}), 2); // \(v\) comp of focus

\nwused{\\{NW3OF8UE-1p0Y9w-1}}\nwidentuses{\\{{\nwixident{u}}{u}}\\{{\nwixident{v}}{v}}}\nwindexuse{\nwixident{u}}{u}{NW3OF8UE-2gOisp-1}\nwindexuse{\nwixident{v}}{v}{NW3OF8UE-2gOisp-1}\nwendcode{}\nwbegindocs{129}Properties of parabolas are printed to {\Tt{}\Rm{}{\it{}stdout}\nwendquote} in the form:
\begin{webcode}Parab (A/ 7/ 0.368); vert=( 1.140, -2.299); l= 0.2500; second vert=(-1.140, -2.299); l=-0.2500\end{webcode}
The two vertexes correspond to two Cayley transformations \(P_e\)
defined by {\Tt{}\Rm{}{\it{}C}\nwendquote} and \(P_h\) defined by {\Tt{}\Rm{}{\it{}C1}\nwendquote},
see~\cite[\S~\ref{E-sec:unit-circles}]{Kisil05a}.
\nwenddocs{}\nwbegincode{130}\sublabel{NW3OF8UE-2qDOKC-1}\nwmargintag{{\nwtagstyle{}\subpageref{NW3OF8UE-2qDOKC-1}}}\moddef{Check parabolas~{\nwtagstyle{}\subpageref{NW3OF8UE-2qDOKC-1}}}\endmoddef\Rm{}\nwstartdeflinemarkup\nwusesondefline{\\{NW3OF8UE-3Fz9v4-2}}\nwprevnextdefs{\relax}{NW3OF8UE-2qDOKC-2}\nwenddeflinemarkup
{\bf{}if} (({\it{}metric} \begin{math}\equiv\end{math} {\it{}parabolic}) \begin{math}\wedge\end{math} ({\it{}subgroup} \begin{math}\neq\end{math} {\it{}subgroup\_K}))
    {\bf{}try} {\nwlbrace}
        {\it{}printf}({\tt{}"{\char92}nParab (
               &{\it{}sgroup}[{\it{}subgroup}], {\it{}vi},  {\it{}vval}, {\it{}output\_focal}(0));
        {\it{}printf}({\tt{}"; second vert=(
\nwalsodefined{\\{NW3OF8UE-2qDOKC-2}}\nwused{\\{NW3OF8UE-3Fz9v4-2}}\nwidentuses{\\{{\nwixident{metric}}{metric}}\\{{\nwixident{output{\_}focal}}{output:unfocal}}\\{{\nwixident{parabolic}}{parabolic}}\\{{\nwixident{sgroup}}{sgroup}}\\{{\nwixident{subgroup}}{subgroup}}\\{{\nwixident{subgroup{\_}K}}{subgroup:unK}}}\nwindexuse{\nwixident{metric}}{metric}{NW3OF8UE-2qDOKC-1}\nwindexuse{\nwixident{output{\_}focal}}{output:unfocal}{NW3OF8UE-2qDOKC-1}\nwindexuse{\nwixident{parabolic}}{parabolic}{NW3OF8UE-2qDOKC-1}\nwindexuse{\nwixident{sgroup}}{sgroup}{NW3OF8UE-2qDOKC-1}\nwindexuse{\nwixident{subgroup}}{subgroup}{NW3OF8UE-2qDOKC-1}\nwindexuse{\nwixident{subgroup{\_}K}}{subgroup:unK}{NW3OF8UE-2qDOKC-1}\nwendcode{}\nwbegindocs{131} In the case of {\Tt{}\Rm{}{\it{}subgroup\_A}\nwendquote} an additional line
\begin{webcode}
  Check vertices: -1 and -1
\end{webcode}
is printed. It confirms that vertexes of the orbits under the Cayley
transform do belong to the parabolas \(v=\pm v^2-1\), as stated in~\cite[\ref{E-le:parabolic-disk}]{Kisil05a}.
\nwenddocs{}\nwbegincode{132}\sublabel{NW3OF8UE-2qDOKC-2}\nwmargintag{{\nwtagstyle{}\subpageref{NW3OF8UE-2qDOKC-2}}}\moddef{Check parabolas~{\nwtagstyle{}\subpageref{NW3OF8UE-2qDOKC-1}}}\plusendmoddef\Rm{}\nwstartdeflinemarkup\nwusesondefline{\\{NW3OF8UE-3Fz9v4-2}}\nwprevnextdefs{NW3OF8UE-2qDOKC-1}{\relax}\nwenddeflinemarkup
        {\bf{}if} ({\it{}subgroup} \begin{math}\equiv\end{math} {\it{}subgroup\_A})
            {\it{}cout} \begin{math}\ll\end{math} {\tt{}"{\char92}nCheck vertices: "}
                 \begin{math}\ll\end{math} {\it{}ex\_to}\begin{math}<\end{math}{\bf{}numeric}\begin{math}>\end{math}({\it{}focal\_v}.{\it{}subs}({\it{}soln}[0]) + {\it{}pow}({\it{}focal\_u}.{\it{}subs}({\it{}soln}[0]), 2).{\it{}evalf}()).{\it{}to\_double}()
                 \begin{math}\ll\end{math} {\tt{}" and "}
                 \begin{math}\ll\end{math} {\it{}ex\_to}\begin{math}<\end{math}{\bf{}numeric}\begin{math}>\end{math}({\it{}focal\_v}.{\it{}subs}({\it{}soln}[1]) + {\it{}pow}({\it{}focal\_u}.{\it{}subs}({\it{}soln}[1]), 2).{\it{}evalf}()).{\it{}to\_double}();
    {\nwrbrace} {\bf{}catch}  ({\it{}exception} &{\it{}p}) {\nwlbrace}
        {\it{}printf}({\tt{}"{\char92}nParab (
               &{\it{}sgroup}[{\it{}subgroup}], {\it{}vi},  {\it{}vval});
    {\nwrbrace}

\nwused{\\{NW3OF8UE-3Fz9v4-2}}\nwidentuses{\\{{\nwixident{catch}}{catch}}\\{{\nwixident{numeric}}{numeric}}\\{{\nwixident{sgroup}}{sgroup}}\\{{\nwixident{subgroup}}{subgroup}}\\{{\nwixident{subgroup{\_}A}}{subgroup:unA}}}\nwindexuse{\nwixident{catch}}{catch}{NW3OF8UE-2qDOKC-2}\nwindexuse{\nwixident{numeric}}{numeric}{NW3OF8UE-2qDOKC-2}\nwindexuse{\nwixident{sgroup}}{sgroup}{NW3OF8UE-2qDOKC-2}\nwindexuse{\nwixident{subgroup}}{subgroup}{NW3OF8UE-2qDOKC-2}\nwindexuse{\nwixident{subgroup{\_}A}}{subgroup:unA}{NW3OF8UE-2qDOKC-2}\nwendcode{}

\nwixlogsorted{c}{{*}{NW3OF8UE-1p0Y9w-1}{\nwixd{NW3OF8UE-1p0Y9w-1}\nwixd{NW3OF8UE-1p0Y9w-2}\nwixd{NW3OF8UE-1p0Y9w-3}\nwixd{NW3OF8UE-1p0Y9w-4}}}%
\nwixlogsorted{c}{{Build future-past transition}{NW3OF8UE-2KWnYy-1}{\nwixu{NW3OF8UE-1p0Y9w-4}\nwixd{NW3OF8UE-2KWnYy-1}\nwixd{NW3OF8UE-2KWnYy-2}}}%
\nwixlogsorted{c}{{Building orbits}{NW3OF8UE-3Fz9v4-1}{\nwixu{NW3OF8UE-1p0Y9w-3}\nwixd{NW3OF8UE-3Fz9v4-1}\nwixd{NW3OF8UE-3Fz9v4-2}}}%
\nwixlogsorted{c}{{Building transverses}{NW3OF8UE-38hDUf-1}{\nwixu{NW3OF8UE-1p0Y9w-3}\nwixd{NW3OF8UE-38hDUf-1}\nwixd{NW3OF8UE-38hDUf-2}}}%
\nwixlogsorted{c}{{C++ variables declaration}{NW3OF8UE-1cUGLU-1}{\nwixu{NW3OF8UE-1p0Y9w-1}\nwixd{NW3OF8UE-1cUGLU-1}}}%
\nwixlogsorted{c}{{Calculation of Moebius transformations}{NW3OF8UE-1OFQOX-1}{\nwixu{NW3OF8UE-1p0Y9w-2}\nwixd{NW3OF8UE-1OFQOX-1}\nwixd{NW3OF8UE-1OFQOX-2}}}%
\nwixlogsorted{c}{{Calculation of vector fields}{NW3OF8UE-4BpIv1-1}{\nwixu{NW3OF8UE-1p0Y9w-2}\nwixd{NW3OF8UE-4BpIv1-1}\nwixd{NW3OF8UE-4BpIv1-2}}}%
\nwixlogsorted{c}{{Check formulas in the paper}{NW3OF8UE-47Sd20-1}{\nwixu{NW3OF8UE-bYD0P-1}\nwixd{NW3OF8UE-47Sd20-1}\nwixd{NW3OF8UE-47Sd20-2}\nwixd{NW3OF8UE-47Sd20-3}\nwixd{NW3OF8UE-47Sd20-4}\nwixd{NW3OF8UE-47Sd20-5}}}%
\nwixlogsorted{c}{{Check parabolas}{NW3OF8UE-2qDOKC-1}{\nwixu{NW3OF8UE-3Fz9v4-2}\nwixd{NW3OF8UE-2qDOKC-1}\nwixd{NW3OF8UE-2qDOKC-2}}}%
\nwixlogsorted{c}{{CiNaC variables declaration}{NW3OF8UE-2Cj0m9-1}{\nwixu{NW3OF8UE-1p0Y9w-1}\nwixd{NW3OF8UE-2Cj0m9-1}\nwixd{NW3OF8UE-2Cj0m9-2}\nwixd{NW3OF8UE-2Cj0m9-3}\nwixd{NW3OF8UE-2Cj0m9-4}}}%
\nwixlogsorted{c}{{Close all curves}{NW3OF8UE-Ix3jQ-1}{\nwixu{NW3OF8UE-3Fz9v4-2}\nwixu{NW3OF8UE-38hDUf-2}\nwixd{NW3OF8UE-Ix3jQ-1}}}%
\nwixlogsorted{c}{{Closing all files}{NW3OF8UE-nrvMS-1}{\nwixu{NW3OF8UE-3Fz9v4-2}\nwixd{NW3OF8UE-nrvMS-1}\nwixu{NW3OF8UE-38hDUf-2}}}%
\nwixlogsorted{c}{{Define transverse directions}{NW3OF8UE-4LL0eE-1}{\nwixu{NW3OF8UE-38hDUf-1}\nwixd{NW3OF8UE-4LL0eE-1}\nwixd{NW3OF8UE-4LL0eE-2}}}%
\nwixlogsorted{c}{{Definitions}{NW3OF8UE-4TccJC-1}{\nwixu{NW3OF8UE-1p0Y9w-1}\nwixd{NW3OF8UE-4TccJC-1}\nwixd{NW3OF8UE-4TccJC-2}\nwixd{NW3OF8UE-4TccJC-3}\nwixd{NW3OF8UE-4TccJC-4}\nwixd{NW3OF8UE-4TccJC-5}\nwixd{NW3OF8UE-4TccJC-6}\nwixd{NW3OF8UE-4TccJC-7}\nwixd{NW3OF8UE-4TccJC-8}\nwixd{NW3OF8UE-4TccJC-9}\nwixd{NW3OF8UE-4TccJC-A}}}%
\nwixlogsorted{c}{{Drawing arrows}{NW3OF8UE-3M9GDT-1}{\nwixu{NW3OF8UE-1p0Y9w-3}\nwixd{NW3OF8UE-3M9GDT-1}}}%
\nwixlogsorted{c}{{Generating one entry}{NW3OF8UE-4P0NfT-1}{\nwixu{NW3OF8UE-bYD0P-1}\nwixu{NW3OF8UE-38hDUf-2}\nwixd{NW3OF8UE-4P0NfT-1}\nwixd{NW3OF8UE-4P0NfT-2}\nwixd{NW3OF8UE-4P0NfT-3}\nwixd{NW3OF8UE-4P0NfT-4}}}%
\nwixlogsorted{c}{{Global items}{NW3OF8UE-4EE31V-1}{\nwixu{NW3OF8UE-1p0Y9w-1}\nwixd{NW3OF8UE-4EE31V-1}\nwixd{NW3OF8UE-4EE31V-2}}}%
\nwixlogsorted{c}{{Includes}{NW3OF8UE-ZKEBO-1}{\nwixu{NW3OF8UE-1p0Y9w-1}\nwixd{NW3OF8UE-ZKEBO-1}}}%
\nwixlogsorted{c}{{Initialisation of coordinates}{NW3OF8UE-2mZmnQ-1}{\nwixd{NW3OF8UE-2mZmnQ-1}\nwixu{NW3OF8UE-3Fz9v4-2}\nwixu{NW3OF8UE-38hDUf-2}}}%
\nwixlogsorted{c}{{Initialise Clifford numbers}{NW3OF8UE-2xd863-1}{\nwixu{NW3OF8UE-1p0Y9w-2}\nwixd{NW3OF8UE-2xd863-1}\nwixd{NW3OF8UE-2xd863-2}\nwixd{NW3OF8UE-2xd863-3}\nwixd{NW3OF8UE-2xd863-4}\nwixu{NW3OF8UE-2KWnYy-1}}}%
\nwixlogsorted{c}{{Iteration over a curve}{NW3OF8UE-3DeWAC-1}{\nwixu{NW3OF8UE-2KWnYy-2}\nwixd{NW3OF8UE-3DeWAC-1}}}%
\nwixlogsorted{c}{{Nodes iterations}{NW3OF8UE-bYD0P-1}{\nwixu{NW3OF8UE-3Fz9v4-2}\nwixd{NW3OF8UE-bYD0P-1}}}%
\nwixlogsorted{c}{{Parabola parameters}{NW3OF8UE-2gOisp-1}{\nwixu{NW3OF8UE-1p0Y9w-1}\nwixd{NW3OF8UE-2gOisp-1}}}%
\nwixlogsorted{c}{{Pictures tuning}{NW3OF8UE-44zkoV-1}{\nwixu{NW3OF8UE-1p0Y9w-1}\nwixd{NW3OF8UE-44zkoV-1}}}%
\nwixlogsorted{c}{{Producing Cayley transform of the orbit}{NW3OF8UE-D1caK-1}{\nwixu{NW3OF8UE-bYD0P-1}\nwixu{NW3OF8UE-38hDUf-2}\nwixd{NW3OF8UE-D1caK-1}\nwixd{NW3OF8UE-D1caK-2}}}%
\nwixlogsorted{i}{{\nwixident{calc{\_}par{\_}focal}}{calc:unpar:unfocal}}%
\nwixlogsorted{i}{{\nwixident{catch}}{catch}}%
\nwixlogsorted{i}{{\nwixident{catch{\_}handle}}{catch:unhandle}}%
\nwixlogsorted{i}{{\nwixident{close{\_}curve}}{close:uncurve}}%
\nwixlogsorted{i}{{\nwixident{color{\_}name}}{color:unname}}%
\nwixlogsorted{i}{{\nwixident{curves}}{curves}}%
\nwixlogsorted{i}{{\nwixident{elliptic}}{elliptic}}%
\nwixlogsorted{i}{{\nwixident{exp{\_}scale}}{exp:unscale}}%
\nwixlogsorted{i}{{\nwixident{fileout}}{fileout}}%
\nwixlogsorted{i}{{\nwixident{get{\_}components}}{get:uncomponents}}%
\nwixlogsorted{i}{{\nwixident{grey}}{grey}}%
\nwixlogsorted{i}{{\nwixident{hyperbolic}}{hyperbolic}}%
\nwixlogsorted{i}{{\nwixident{if{\_}in{\_}limits}}{if:unin:unlimits}}%
\nwixlogsorted{i}{{\nwixident{init{\_}coord}}{init:uncoord}}%
\nwixlogsorted{i}{{\nwixident{main}}{main}}%
\nwixlogsorted{i}{{\nwixident{metric}}{metric}}%
\nwixlogsorted{i}{{\nwixident{name}}{name}}%
\nwixlogsorted{i}{{\nwixident{numeric}}{numeric}}%
\nwixlogsorted{i}{{\nwixident{openfile}}{openfile}}%
\nwixlogsorted{i}{{\nwixident{output{\_}focal}}{output:unfocal}}%
\nwixlogsorted{i}{{\nwixident{parabolic}}{parabolic}}%
\nwixlogsorted{i}{{\nwixident{put{\_}draw}}{put:undraw}}%
\nwixlogsorted{i}{{\nwixident{put{\_}point}}{put:unpoint}}%
\nwixlogsorted{i}{{\nwixident{renew{\_}curve}}{renew:uncurve}}%
\nwixlogsorted{i}{{\nwixident{S}}{S}}%
\nwixlogsorted{i}{{\nwixident{sgroup}}{sgroup}}%
\nwixlogsorted{i}{{\nwixident{subgroup}}{subgroup}}%
\nwixlogsorted{i}{{\nwixident{subgroup{\_}A}}{subgroup:unA}}%
\nwixlogsorted{i}{{\nwixident{subgroup{\_}K}}{subgroup:unK}}%
\nwixlogsorted{i}{{\nwixident{subgroup{\_}N}}{subgroup:unN}}%
\nwixlogsorted{i}{{\nwixident{transverse{\_}dir}}{transverse:undir}}%
\nwixlogsorted{i}{{\nwixident{u}}{u}}%
\nwixlogsorted{i}{{\nwixident{ulim}}{ulim}}%
\nwixlogsorted{i}{{\nwixident{upos}}{upos}}%
\nwixlogsorted{i}{{\nwixident{v}}{v}}%
\nwixlogsorted{i}{{\nwixident{vilimits}}{vilimits}}%
\nwbegindocs{133}\nwdocspar

\section{How to Get the Code}
\label{sec:how-get-code}

\begin{enumerate}
\item Get the  \href{http://www.arxiv.org/abs/cs.MS/0410044}{\LaTeX\
    source of this paper}~\cite{Kisil04c} from the
  \href{http://www.arxiv.org/abs/cs.MS/0410044}{arXiv.org}.
\item Run the source through \LaTeX. Three new files (\NoWEB, \CPP\ and
  \MetaPost\ sources) will be created in the current directory.
\item Use it on your own risk under the GNU General Public License
  \cite{GNUGPL}.

\end{enumerate}

\appendix

\section{Textual Output of the Program}
Here is the complete textual output of the program.

\label{sec:append-text-outp}
\begin{webcode}
\input{parabolic0-out.d}
\end{webcode}

\begin{figure}[ht]
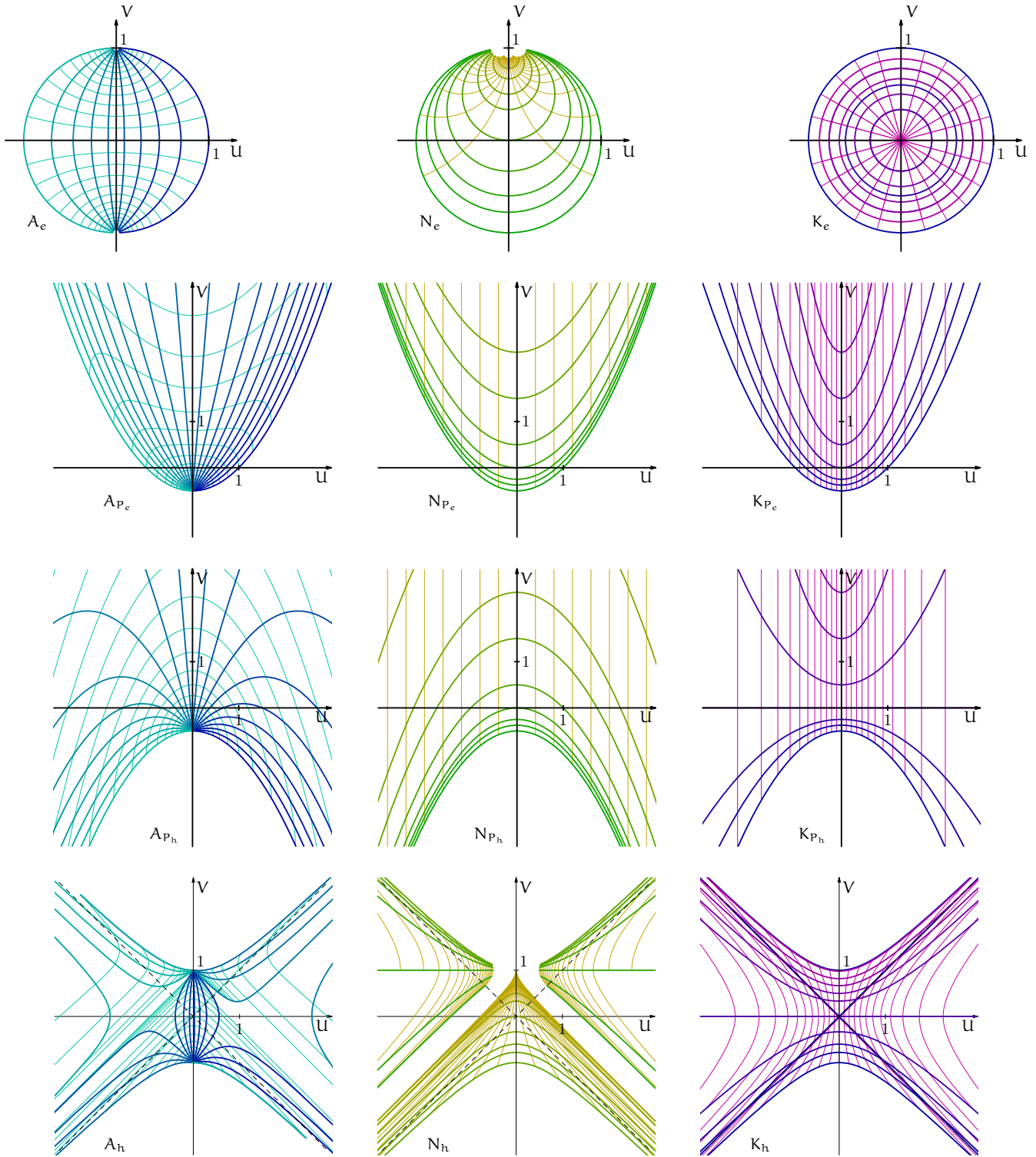

\includegraphics[scale=.8]{parabolic0.40}\hfill
\includegraphics[scale=.8]{parabolic0.41}\hfill
\includegraphics[scale=.8]{parabolic0.42}\\[4mm]
\includegraphics[scale=.8]{parabolic0.50}\qquad
\includegraphics[scale=.8]{parabolic0.51}\qquad
\includegraphics[scale=.8]{parabolic0.52}\\[4mm]
\includegraphics[scale=.8]{parabolic0.53}\qquad
\includegraphics[scale=.8]{parabolic0.54}\qquad
\includegraphics[scale=.8]{parabolic0.55}\\[4mm]
\includegraphics[scale=.8]{parabolic0.60}\qquad
\includegraphics[scale=.8]{parabolic0.61}\qquad
\includegraphics[scale=.8]{parabolic0.62}
\caption{The elliptic, parabolic and hyperbolic unit disks.}
  \label{fig:unit-disks}
\end{figure}

\section{A Sample of Graphics Generated by the Program}
\label{sec:graph-gener-progr}
A sample of graphics produced by the program and post-processed
with \MetaPost is shown on Figure~\ref{fig:unit-disks}. Some more
examples can be found in~\cite{Kisil05a}.

\section{Index of Identifiers}
\label{sec:index-identifiers}

\nowebindex
\bibliographystyle{plain}
\bibliography{arare,aclifford,abbrevmr,akisil,analyse,aphysics}
\end{document}